\documentclass[twocolumn]{autart}    

\usepackage{graphicx}          

\usepackage[T1]{fontenc}
\usepackage[utf8]{inputenc}
\usepackage{ctable}
\usepackage{tikz}
\usetikzlibrary{arrows,automata}
\usepackage{graphicx}
\usepackage{subcaption}
\usepackage{amsfonts}
\usepackage{amssymb}
\usepackage{amsmath}
\usepackage{tabulary}
\usepackage{multirow}

\usepackage[noprefix]{nomencl}
\makenomenclature

\usepackage{url}
\usepackage{acronym}
\usepackage{enumerate}
\usepackage{breakurl} 
\newtheorem{theorem}{\bf Theorem}[section]
\usepackage[noend]{algpseudocode}
\usepackage{soul}
\usepackage{algorithm}
\newcommand{\algrule}[1][.2pt]{\par\vskip.5\baselineskip\hrule height #1\par\vskip.5\baselineskip}
\newtheorem{lemma}[theorem]{\bf Lemma}
\newtheorem{proposition}[theorem]{\bf Proposition}

\newtheorem{definition}{\bf Definition}[section]

\begin{document}

\begin{frontmatter}

\title{Sensor placement for fault location identification in water networks: a minimum test cover approach} 
\author[MIT]{Lina Sela Perelman}\ead{linasela@mit.edu},    
\author[Vanderbilt]{Waseem Abbas}\ead{waseem.abbas@vanderbilt.edu},
\author[Vanderbilt]{Xenofon Koutsoukos}\ead{ Xenofon.Koutsoukos@vanderbilt.edu },               
\author[MIT]{Saurabh Amin}\ead{amins@mit.edu}  

\address[MIT]{Massachusetts Institute of Technology}  
\address[Vanderbilt]{Vanderbilt University}             
\vspace{-0.35cm}

\begin{abstract}                          
This paper focuses on the optimal sensor placement problem for the identification of pipe failure locations in large-scale urban water systems. The problem involves selecting the minimum number of sensors such that every pipe failure can be uniquely localized. This problem can be viewed as a minimum test cover (MTC) problem, which is NP-hard.  We consider two approaches to obtain approximate solutions to this problem. In the first approach, we transform the MTC problem to a minimum set cover (MSC) problem and use the greedy algorithm that exploits the submodularity property of the MSC problem to compute the solution to the MTC problem. 
In the second approach, we develop a new \textit{augmented greedy} algorithm for solving the MTC problem. This approach does not require the transformation of the MTC to MSC. Our augmented greedy algorithm provides in a significant computational improvement while guaranteeing the same approximation ratio as the first approach. We propose several metrics to evaluate the performance of the sensor placement designs. Finally, we present detailed computational experiments for a number of real water distribution networks.\vspace{-0.35cm}
\end{abstract}
\end{frontmatter}

\section{Introduction}\vspace{-0.35cm}
Infrastructure deterioration, demand-supply uncertainty, and risk of disruptions pose new challenges in maintaining modern infrastructures. Resilient urban infrastructures including water distribution systems, transportation networks, and electric grids are crucial for societal well-being. \textit{Smart} infrastructure operation driven by sensing and actuation technologies have been identified as one of the primary solutions towards resilient urban systems \cite{urbaninfra:edit,6855316}.
Through a network of sensors, an individual fault or correlated failures in a system component can be detected and localized, and restorative actions can be executed in response to these faults.
Whereas network observability for a given sensing capability has been widely studied in the context of fault detection, sensor placement for fault isolability, i.e. the ability to distinguish between faults, has not been a commonly studied problem, especially in the context of pipe bursts in water distribution networks. \vspace{-0.35cm}

The goal of this work is to \textit{design a sensor placement configuration for identification of pipe failure locations by using the minimum number of sensors.} The underlying idea behind our approach is to ensure that the sensor placement results in a collective output that is \textit{unique} for each failure event. 
Specifically, our main contributions are as follows, we: \vspace{-0.25cm}
\begin{itemize}
\item[--] Define the \textit{localization} of pipe bursts as the design objective of a sensor network configuration, and using ideas from combinatorial optimization, we formulate the fault location identification problem as a \textit{minimum test cover} (MTC) problem;
\item[--]Develop a computationally efficient \textit{augmented greedy} algorithm to solve the minimum test cover problem (resp. identification problem), which is significantly faster in comparison to the previous approaches and therefore, scalable to large-scale networks; and
\item[--] Test and evaluate our sensor placement approach on a batch of real-networks of various sizes and parameters using practically relevant performance measures.  
\end{itemize}\vspace{-0.15cm}

Our paper is motivated by the need to consider localization of pipe bursts in the deployment phase of new sensing technologies, since this consideration can significantly reduce the response time and overall costs of fault localization to the distribution utilities.
We base our work on the use of low-cost, high-rate online sensors measuring water pressure for remote detection of pipe burst using data mining techniques. Real-world examples are the PIPENET in Boston, MA, US \cite{Stoianov:2007:PWS:1236360.1236396} and the WaterWise in Singapore \cite{allen:smart}. 
The sensor placement problem is not unique to the water sector and can be found in many engineering applications for system operation. We discuss some of the related work in Section \ref{sec:review}.
\vspace{-0.35cm}

In Section \ref{sec:problem}, we present the network and the sensing models and formulate the detection and identification problems as the minimum set cover (MSC) and minimum test cover (MTC) problems, respectively. A key aspect of the MTC problem formulation is the choice of the objective function, which is to select the minimum number of \textit{tests} from a collection of tests such that every event can be uniquely classified in one of the given \textit{categories} based on selected tests' outcomes \cite{Moret1}. In our setup, the set of outcomes of tests comprise of the output vector from sensors, events are pipe failures, and classification categories are the possible locations of the failed pipes. 
In Section \ref{sec:oldsol}, we present a solution approach as in \cite{Halldorsson:2001:AMT:647911.740635,svard2013realizability}, in which the MTC is first transformed to the MSC and then solved using the greedy approximation \cite{Minoux1}. \vspace{-0.25cm}

In Section \ref{sec:New_Algo} we present an \textit{augmented greedy} algorithm for solving the MTC  that does not require the complete transformation of the MTC to the equivalent MSC, and directly computes the objective function in a greedy fashion. This algorithm is much faster than the standard greedy approach and considerably improves the scalability of our approach.
In Sections \ref{sec:applications} and \ref{sec:app2}, we demonstrate our approach using a benchmark and a batch of twelve real water distribution networks of various sizes and specifications. We suggest four metrics to evaluate the performance of the design including detection, identification, and localization scores. Although we demonstrate our results in the context of water networks, our algorithm provides an improved solution to the generic test cover problem.
Section \ref{sec:conclusions} summarizes our work and proposes future extensions.

\section{Problem formulation}\label{sec:problem}\vspace{-0.35cm}
Consider the problem of placing online sensors measuring hydraulic pressures with high frequency such that the identification of pipe failure locations is maximized. Based on the number of pipes where link failures (i.e., pipe bursts) can happen, we consider $n$ link failures as a set of failure events, denoted by $\mathcal{L}  = ~ \{ \ell_1, \dots, \ell_n \}$. For the ease of presentation and without the loss of generality, let $\ell_j$ denote the failure event at the $j^{th}$ pipe. Moreover, we define a set of  sensors that can be placed at $m$ nodes of the network as $\mathcal{S}= \{ S_1, \dots, S_m \}$. Here, $S_i$ denotes the location of the $i^{th}$ sensor. 
The outputs from sensors, which are based on the change in pressure induced by the failure event, are denoted by $\mathbf{y}_{{\mathcal{S}}}$.

\subsection{Network dynamics and sensing model}\vspace{-0.35cm}
A water distribution network can be represented by a graph comprising nodes (supply and demand) connected by links (pipes, valves, and pumps). 
Physical failures of the infrastructure, such as pipe bursts, cause a disturbance in the flow, which moves through the system as a pressure wave known as \textit{water hammer}, or \textit{surge} with very high velocity, varying typically in the range of $600 - 1500 [\frac{m}{s}]$ \cite{Dalius}. This implies that the steady state analysis employed by traditional methods such as supervisory control and data acquisition (SCADA) systems are inadequate and that the transient system dynamics between the initial and the final steady state conditions need to be considered. \vspace{-0.25cm}

The transient system state can be typically described by mass and momentum partial differential equations \cite{wylie1993fluid}.
The method of characteristics (MOC) is a numerical technique typically used to approximate the solution of the hydraulic transients. The MOC transforms the partial differential equations into ordinary differential equations that evolve along specific characteristic lines of the numerical grid, which are solved explicitly to compute the head and flow, $h_{i,t+1},q_{i,t+1}$, at new point in time and space. Here, $t$ and $i$ indicate the discrete points of the numerical grid. For a given pipe, the two characteristic equations describing the hydraulic transients are formulated as \cite{Dalius}:
\begin{align} 
 h_{i,t+1} &= \frac{1}{2} \big[ h_{i-1,t} + h_{i+1,t} + b \left( q_{i-1,t} - q_{i+1,t} \right) \nonumber \\
 &\qquad {} + r \left( q_{i+1,t} |q_{i+1,t}| - q_{i-1,t} |q_{i-1,t}| \right) \big] \label{eq:3a}\\
q_{i,t+1} &= \frac{1}{b} \big[ h_{i,t+1} -h_{i+1,t} + q_{i+1,t} - r |q_{i+1,t}| \big],  \label{eq:3b}
\end{align} 
where $r$ is the resistance coefficient associated with the steady state, and $b$ is the impedance coefficient associated with the transient state. For $b = 0$ the set of equations \eqref{eq:3a},\eqref{eq:3b} is reduced to the steady state, where the head loss along a pipe occurs only due to friction \cite{todini}. Additional information describing transient dynamics can be found in the supporting information (SI) \cite{perelman2015}. \vspace{-0.25cm} 

The effect of a pipe burst at location $i$ can be translated into boundary conditions using the orifice head-flow relation \cite{wylie1993fluid}. Before the burst occurs, the cross-section area of the orifice is equal to zero and it increases during a burst, hence we can expect a sudden change in the hydraulic head. The relationship between the head and the pressure, measured by the sensors at location $i$, is related to the elevation of the sensor location. If $z_i$ is the elevation, and $p_{i,t}$ is the pressure at location $i$ at any given time $t$, then
$p_{i,t} = \left( h_{i,t} - z_i \right)\rho g$, 
where $g$ is the gravitational acceleration $[\frac{m}{sec^2}]$ and $\rho$ is water density $[\frac{kg}{m^3}]$.  
Hence, the disturbance caused by a pipe burst that reaches the sensor location can be detected by sensing the hydraulic pressure. Similar approaches have been suggested in \cite{6901220}.  \vspace{-0.25cm}

%

The disturbance caused by the pipe burst quickly dissipates with the distance between the burst event $\ell_j$ and the location of the sensor $S_i$. 
For the purpose of sensor placement, we are interested in obtaining the sensor's output as a result of some event $\ell_j$. 
Let $y_{S_i}(t,\ell_j) \in \{0,1\} $ be a discrete state (output) of the sensor $S_i$ at time $t$, where 1 represents a possible detected event and 0 represents otherwise. 
Let $\xi$ be a function characterizing the distance between the expected pressure (i.e., when there is no pipe burst), denoted by $\hat{p}_{i,t}$, and the measured pressure, denoted by $p_{i,t}$. %
The sensor output can then be formulated as:\vspace{-0.15cm}
\begin{equation} \label{eq:3}
y_{S_i}(t,\ell_j) =
\left\{
\begin{array}{lcl}
1 & \;\;\;\text{if}\;\; \xi \left( p_{i,t} - \hat{p}_{i,t}\right) \geq \varepsilon  \\
0 & \;\;\;\text{otherwise}\\
\end{array}
\right.
\end{equation}
where $\varepsilon$ is a threshold value. A simple detection model would be where the sensor $S_i$ indicates an event if the change in the pressure is above some threshold value $\varepsilon $. 
We note here that when the failure event $\ell_j$ occurs during a given time period, then the output of $S_i$ will be $1$ (or $0$) independent of the time of the event $\ell_j$. 
Hence, we can neglect the time dependency of the sensor output to detect the event and can restate the output of the sensor as:\vspace{-0.25cm}
\begin{equation}
\mathbf{y}_{S_i}(\ell_j) = 
\left\{
\begin{array}{lcl}
 1 & \;\;\;\text{if}\;\;y_{S_i}(t,\ell_j) = 1,\;\;\text{for any }t>0 \\
 0 & \;\;\;\text{otherwise}\\
\end{array}
\right.
\end{equation}
Let $\mathbf{y}_{\mathcal{S}}(\ell_j) = \left[ \mathbf{y}_{S_1}(\ell_j), \cdots,  \mathbf{y}_{S_m}(\ell_j) \right]$ be the fault signature \cite{1335513} of the failure event $\ell_j$ represented by a boolean vector of the outputs of sensors in the set $\mathcal{S}$. \vspace{-0.25cm}

Consequently, for a sensor set $\mathcal{S}$ and the set of events $\mathcal{L}$, we can instantiate a boolean matrix of dimensions $\lvert \mathcal{L} \rvert\times \lvert {\mathcal{S}} \rvert$ called the \textit{influence matrix} and denoted by $\mathcal{M}$. The $j^{th}$ row of $\mathcal{M}$ consists of sensors' outputs in response to the event $\ell_j$, i.e., $\mathbf{y}_\mathcal{S}(\ell_j)$. Similarly, $\mathcal{M}_{ij}=1$ indicates that a sensor $S_i$ detected the failure at link $\ell_j$, and $\mathcal{M}_{ij}=0$ means otherwise. Each row of the influence matrix $\mathcal{M}$ is analogous to the notion of fault signature in the model-based fault diagnosis systems literature \cite{1335513}. 
\vspace{-0.25cm}
\begin{equation}
\label{eq:sensing_model}
\mathcal{M}\left(\mathcal{L},{\mathcal{S}}\right) =
\left[
\begin{array}{c}
\mathbf{y}_{{\mathcal{S}}}(\ell_1)\\
\mathbf{y}_{\mathcal{S}}(\ell_2)\\
\vdots\\
\mathbf{y}_{\mathcal{S}}(\ell_n)\\
\end{array}
\right]
\end{equation}
Furthermore, for the set of link failures $\mathcal{L}$, and the set of all possible sensor locations $\mathcal{S}$, let $C_i\subseteq \mathcal{L}$ be the set of link failure events detected by the sensor $S_i$, i.e., $C_i=\{\ell_j\in\mathcal{L}| \; \mathbf{y}_{S_i}(\ell_j)=1\}$. If $\mathcal{C}$ is a collection of all such $C_i$'s, i.e., $\mathcal{C}=\{C_i:\;\forall i\}$, then for a given subset of sensors $S\subseteq\mathcal{S}$, we define $\mathcal{C}_S\subseteq \mathcal{C}$ as a set of subsets of failure events, where a subset corresponds to a sensor in $S$ that detects the failure events in that subset, i.e., $\mathcal{C}_S=\{C_i:\; S_i\in S\}$. \vspace{-0.25cm}
\begin{exmp}[Sensing model] \label{ex:1}
\begin{figure}[ht]
\centering
\includegraphics[scale = 0.5]{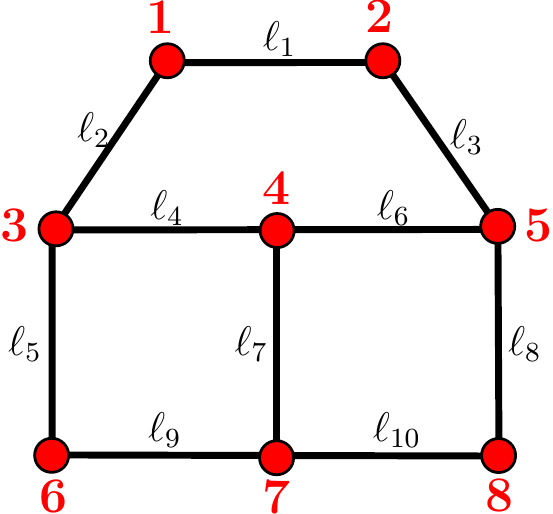}
\caption{Illustrative example layout}
\label{fig:3}
\end{figure}
To illustrate the network dynamics, consider a small network having  $8$ nodes connected by $10$ links as shown the Figure \ref{fig:3}. A pipe burst event is simulated in the middle of pipe $\ell_1$ and system response at network nodes is recorded. For the ease of notations, we designate the failure events as pipes' ids, $\ell_j$. The transient simulations were computed using the HAMMER software \cite{Bentley}. 
Figure \ref{fig:4} shows simulated pressure heads and boolean outputs $\mathbf{y}_{S}$, for sensors located at nodes 2 and 4. 
Thus for $S = \{S_2,S_4\}$ the sensors' state is $\mathbf{y}_{S} (\ell_1)  = [1,0]$. If sensors are placed at all nodes of the network, then the sensors' state in the case of failure at $\ell_1$ is $\mathbf{y}_{\mathcal{S}} (\ell_1) = [1,1,1,0,1,0,0,0]$, $\mathbf{y}_{\mathcal{S}} (\ell_2) = [1,1,1,1,0,1,0,0]$, and so on. 
\begin{figure}
        \centering       
   \includegraphics[trim = 10mm 0mm 30mm 130mm, clip, scale = 0.35]{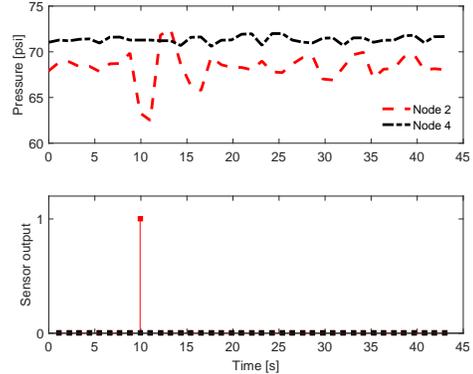}
       \caption{Failure event generated in pipe $\ell_1$ in the small example -- pressure head $[m]$ and outputs of sensors $S_2,S_4$.}  \label{fig:4}
\end{figure}
The corresponding influence matrix is \vspace{-0.25cm}
\scriptsize
$$\mathcal{M}( \mathcal{L},\mathcal{S}) = 
\bordermatrix{
& S_1& S_2& S_3& S_4& S_5& S_6& S_7& S_8\cr
\ell_1&1 & 1 & 1 & 0 & 1 & 0 & 0 & 0 \cr 
\ell_2&1 & 1 & 1 & 1 & 0 & 1 & 0 & 0 \cr
\ell_3&1 & 1 & 0 & 1 & 1 & 0 & 0 & 1 \cr 
\ell_4&1 & 0 & 1 & 1 & 1 & 1 & 1 & 0 \cr
\ell_5&1 & 0 & 1 & 1 & 0 & 1 & 1 & 0 \cr
\ell_6&0 & 1 & 1 & 1 & 1 & 0 & 1 & 1 \cr 
\ell_7&0 & 0 & 1 & 1 & 1 & 1 & 1 & 1 \cr
\ell_8&0 & 1 & 0 & 1 & 1 & 0 & 1 & 1 \cr
\ell_9&0 & 0 & 1 & 1 & 0 & 1 & 1 & 1 \cr
\ell_{10}& 0 & 0 & 0 & 1 & 1 & 1 & 1 & 1}.
$$
\end{exmp} \vspace{-0.35cm}
Next, we formulate the detection and identification problems as the minimum set and test cover problems, respectively.

\subsection{Detection as MSC} \label{sec:dmsc}\vspace{-0.35cm}
For the set of events $\mathcal{L}$ and the set of sensors $\mathcal{S}$, we define a \textit{detectable event} as the one for which there exists at least one sensor in $\mathcal{S}$ that detects the event. The \textit{detection problem} is to select the minimum number of sensors $S\subseteq\mathcal{S}$, such that when a detectable event occurs, at least one sensor in $S$ detects the event. For a given subset of sensors $S$, we define the \textit{detection function}, denoted by $f_D$, as follows:
\vspace{-0.35cm}
\begin{equation}\label{eq:6}
f_D(\mathcal{C}_S) = \left\lvert \bigcup\limits_{C_i\in\mathcal{C}_S}C_i\right\rvert.
\end{equation} 
The detection function in \eqref{eq:6} gives the number of link failures in $\mathcal{L}$ that can be detected by the sensors in $S$. The detection problem is to select a subset of sensors $S\subseteq\mathcal{S}$ with the minimum cardinality such that all detectable events are detected, i.e. $f_D(\mathcal{C}_S) = f_D(\mathcal{C}_\mathcal{S})$. The detection performance of a subset of sensors $S$ is defined as the \textit{normalized detection score}, $I_D(S)$ and is computed as $f_D(\mathcal{C}_S)/|\mathcal{L}|$. 
The detection problem is equivalent to the \textit{minimum set cover} problem, which could be defined as: \vspace{-0.15cm}

\begin{definition} (Minimum set cover (MSC))
\label{def:set_cover}
Let $\mathcal{L}$ be a finite set of elements, and $\mathcal{C}=\{C_i:\;C_i\subseteq\mathcal{L}\}$ be the collection of given subsets of $\mathcal{L}$. The minimum set cover is to find $\mathcal{C}_s\subseteq\mathcal{C}$ with the minimum cardinality such that $\bigcup\limits_{C_i\in\mathcal{C}}C_i= \bigcup\limits_{C_j\in\mathcal{C}_s}C_j$.
\end{definition} \vspace{-0.15cm}

In the above definition, if $\mathcal{L}$ is the set of link failures and $\mathcal{C}$ is the collection of $C_i$'s corresponding to all the available sensors, then a set cover of minimum size $\mathcal{C}_s$, gives the minimum number and locations of sensors that solve the detection problem. 
Thus, we get the following:
\begin{proposition}
The problem of detection of link failures in a network is equivalent to the minimum set cover problem, and a solution to MSC is therefore, a solution to the detection problem.
\end{proposition}

The MSC problem is closely related to the \textit{maximum coverage} problem \cite{vazirani2013approximation}, which emerges when the number of sensors that could be used is limited, i.e., $|S|\le B$. The objective of the maximum coverage problem is to select the sensors such that the number of detectable events is maximized and the constraint $|S|\le B$ is satisfied. In Section \ref{sec:ds} we discuss the \textit{greedy} solution approach, which is very much similar for the MSC and the maximum coverage problems.


\subsection{Identification as MTC}\vspace{-0.35cm}
\label{sec:Ident_as_MTC}
For the identification of link failures, the goal is to \textit{uniquely} detect the events in $\mathcal{L}$, i.e. to distinguish between events using the outputs of sensors. We note that event $\ell_i\in\mathcal{L}$ can be distinguished from event $\ell_j\in\mathcal{L}$, if there exists a sensor in $\mathcal{S}$ that gives different outputs for $\ell_i$ and $\ell_j$. In such a case, we say that the \textit{pair-wise event} $\ell_i,\ell_j$ is \textit{detectable} 
if $\exists S_p\in\mathcal{S}:\;\mathbf{y}_{S_p}(\ell_i)\ne\mathbf{y}_{S_p}(\ell_j)$. In terms of the influence matrix of the network, if a pair-wise event $\ell_i,\ell_j$ is detectable, then there exists a column with different $i$ and $j$ row entries. It follows that an event $\ell_i$ can be uniquely detected if all pair-wise events $\ell_i,\ell_j, \;\forall j\ne i$ are detectable. 

The \textit{identification problem} is now defined as follows: 
\textit{for a given $\mathcal{L}$ and $\mathcal{S}$, the identification problem is to select a subset of sensors $S\subseteq\mathcal{S}$ with the minimum cardinality, such that every detectable pair-wise event can be detected by at least one sensor in $S$.} The \textit{identification function} of $S$,  $f_I(\mathcal{C}_S)$, is the number of pair-wise events that are detected by a subset of sensors $S\subseteq \mathcal{S}$, and will be further discussed in Section \ref{sec:t2s}. 
The identification problem is equivalent to the \textit{minimum test cover} problem, which is defined as follows \cite{Bontriddr1}:\vspace{-0.15cm}

\begin{definition} (Minimum test cover (MTC))
\label{def:test_cover}
Consider a finite set $\mathcal{L}$ and a collection of subsets $\mathcal{C}=\{C_i: \;C_i\subseteq\mathcal{L}\}$. The minimum test cover is to find $\mathcal{C}_t\subseteq\mathcal{C}$ with the minimum cardinality such that if for a pair of elements $\{\ell_u,\ell_v\} \in\mathcal{L}$, there exists ${C}_i\in\mathcal{C}$ that contains either $\ell_u$ or $\ell_v$ but not both, then there exists some ${C}_j\in\mathcal{C}_t$ that also contains either $\ell_u$ or $\ell_v$, but not both.
\end{definition}

The identification problem is to find a subset $\mathcal{C}_t\subseteq\mathcal{C}$ of minimum cardinality, or equivalently the corresponding subset of sensors $S\subseteq\mathcal{S}$, such that if $\mathbf{y}_\mathcal{S}(\ell_j)$ is unique with respect to the set of all sensors $\mathcal{S}$, then $\mathbf{y}_S(\ell_j)$ is also unique with respect to a subset of sensors $S$, which is the MTC problem defined above. Thus, we can state:

\begin{proposition} The problem of identification of link failures in networks is equivalent to the minimum test cover problem, and therefore, a solution to MTC is also a solution to the identification problem.
\end{proposition}

\begin{exmp} [Detection vs. Identification]\label{exmp:2}
Follow-ing example \ref{ex:1}, consider two sensors placed at nodes 2 and 4, $S = \{S_2,S_4\}$. 
For the detection problem, we note that $C_2\cup C_4 = \mathcal{L}$. That is, at least one of the sensors in $S$ has an output $1$ whenever a link fails. Thus, sensors $S_2$ and $S_4$ \textit{cover} (detect) all link failures and solve the detection problem. 
For the identification problem, sensors $2$ and $4$ are not sufficient as they generate only three unique states associated with the 10 events, which makes it impossible to distinguish between all link failures. For example, the state $\{1,0\}$ is uniquely associated with a failure in link $\ell_1$, whereas, the state $\{1,1\}$ can be associated with a failure in any of the links $\ell_2,\ell_3, \ell_6, \text{or } \ell_8$. 
However, for the set of sensors $S^* = \{S_1,S_2,S_3,S_5 \}$, which solves the MTC problem for example \ref{ex:1}, the output is unique for each link failure, i.e. ten distinct indicator vectors, each corresponding to a unique failure event, are obtained.
\end{exmp}

\section{Greedy MTC solution} \label{sec:oldsol} \vspace{-0.35cm}
It is well known that both MSC and MTC are NP-hard problems \cite{Garey1,vazirani2013approximation}. In this section, we first introduce an approximate solution to the MSC, which will be utilized in Section \ref{sec:New_Algo} for constructing a computationally efficient solution of the MTC problem.

\subsection{Detection solution}\label{sec:ds}\vspace{-0.35cm}
MSC has been studied extensively owing to its wide variety of applications in theoretical as well as practical domains. A straight-forward way to solve the MSC is by the \textit{greedy approach}. The greedy approach is to select, in each iteration, a sensor that detects the maximum number of undetected link failures, until all link failures are detected, or no further link failure can be detected by any sensor. In the maximum coverage problem, iterations continue until a given number of sensors are selected. If $n$ is the total number of link failures, $m$ is the total number of sensors, then \textit{greedy} algorithm for the MSC gives the best approximation ratio of $\mathcal{O}(\ln n)$ \cite{Garey1,Lund1}. In fact, if $k$ is the maximum number of link failures that can be detected by any sensor, then the greedy algorithm has an approximation ratio of $\mathcal{O}(\ln k)$, which is the best possible (unless P=NP) \cite{vazirani2013approximation}. In our context, $k$ depends on the network topology and the sensing model as in (\ref{eq:sensing_model}). Similarly, for the maximum coverage problem, the greedy algorithm gives the approximation ratio of $(1-1/e)$, which is again the best possible.

Although the greedy approach gives the best known approximation ratio, its straightforward implementation requires a large number of function (as in (\ref{eq:6})) evaluations. The running time of greedy approach is a function of the number of sensors and events, $\mathcal{O}(mn)$. For large scale systems, in which $n$ and $m$ are very large, this simple greedy approach becomes computationally intractable owing to a large number of function evaluations, even if computing a function is not expensive. However, greedy algorithm can be made faster by reducing the number of function evaluations if the \textit{submodularity} property is satisfied \cite{Minoux1}. Submodular functions can be defined as follows:

\begin{definition} (Submodularity)
\label{def:submodular}
Let $\mathcal{C}$ be a finite set and $f$ be a set function, $f:2^\mathcal{C}\longrightarrow \mathbb{R}$. Moreover, $\mathcal{C}_s\subseteq \mathcal{C}_r\subseteq\mathcal{C}$, and $C_i\in\mathcal{C}\setminus \mathcal{C}_r$, then $f$ is submodular whenever \vspace{-0.25cm}
\begin{equation}
\label{eq:submodular}
f\left(\mathcal{C}_s\cup\{C_i\}\right) - f(\mathcal{C}_s) \ge f\left(\mathcal{C}_r\cup\{C_i\}\right) - f(\mathcal{C}_r)
\end{equation}   
\end{definition} \vspace{-0.35cm}

For the detection problem, this means that as the number of link failures detected by the selected sensors increases, the marginal value of adding a sensor to the cover decreases. 
It can be shown that the function in \eqref{eq:6} is submodular (see \cite{perelman2015}), and the submodularity of $f_D$ can be exploited to obtain the \textit{lazy greedy} algorithm as in \cite{Minoux1}. The basic idea behind the lazy greedy approach is to eliminate the redundant computations in each iteration. This can be further explained as follows: For the $\kappa^{th}$ iteration, let $F_\kappa(C_i)$ denotes the utility of adding a sensor $i$ to the cover, i.e. $f_D(\mathcal{C}_s\cup\{C_i\} - f_D(\mathcal{C}_s)$, then by the submodularity of $f_D$, we know that $F_{\kappa+1}(C_i)\le F_\kappa(C_i)$. Moreover, without the loss of generality, we assume that $F_\kappa(C_1)\ge F_\kappa(C_2)\ge\dots$, then $C_1$ is the greedy choice in the $\kappa^{th}$ iteration. However, in the next iteration, if $F_{\kappa+1}(C_2)\ge F_\kappa(C_3)$, then $F_{\kappa+1}(C_2)\ge F_{\kappa+1}(C_j),$ $\forall j\ge 3$, which means that there is no need to compute $F_{\kappa+1}(C_j),\;\forall j\ge 3$. This saves a large number of potential computations and improves scalability of the solution approach to large scale systems.

\subsection{Identification solution} \label{sec:3.2}\vspace{-0.35cm}
One approach to solve the MTC problem is to first transform it to an equivalent MSC problem  \cite{Bontriddr1}, and then to solve the MSC problem using lazy greedy algorithm, as explained earlier. The greedy approach to solve the MTC yields a $(2\ln n + 1)$ approximation ratio algorithm, which is the best possible \cite{Moret1}. 
A solution of the equivalent MSC is a solution to the original MTC problem. Thus, a straight-forward way to solve the identification problem for link failures is to first obtain an equivalent detection problem, in which each event represents a \textit{pair-wise} link failure, and then utilize the greedy approach to solve the corresponding detection problem. We call this the \textit{transformed lazy greedy (TLG)} and will use it in Section \ref{sec:app} to demonstrate the simulation results.
Next, we summarize the transformation of the MTC to the MSC problem as outlined in \cite{Bontriddr1}.

\subsubsection{Transformation of MTC to MSC} \label{sec:t2s} \vspace{-0.35cm}
Given an instance of the MTC, i.e., $\mathcal{L}$ and $\mathcal{C}=\{C_i\}$, where $C_i\subseteq\mathcal{L}$, we transform the MTC to the MSC by taking the following two steps:
\begin{itemize}
\item \textit{Create a new set of events}: $\mathcal{L}^t= \{\ell^t_{12}, \cdots,\ell^t_{(n-1)n}\}$.
For each unordered pair $\{\ell_i,\ell_j\}$, define a new element $\ell_{ij}^t$; $\mathcal{L}^t$ consists of all such $\ell_{ij}^t$'s.
\item \textit{Create a new sets of sensors' outputs}: $\mathcal{C}^t =\{C_1^t,\cdots,C_m^t\}$, where $C_v^t = 
\{\ell_{ij}^t:\lvert\{\ell_i,\ell_j\}\cap C_v\rvert = 1\}, \;\forall k\in\{1,\cdots,m\}$. In other words, $\ell_{ij}^t\in C_v^t$ if and only if exactly one of $\ell_i$ or $\ell_j$ is in $C_v$. Moreover, for a subset of sensors $S\subseteq\mathcal{S}$, we define $\mathcal{C}^t_S = \{C^t_v: S_v\in S\}$.
\end{itemize}

Hence, we obtain a new identification matrix $\mathcal{M}^t(\mathcal{L}^t,\mathcal{S})$ of dimensions $\dbinom{n}{2}\times m$, in which each row corresponds to a \textit{pair-wise link failure} and each column represents sensor's output. If a specific row in $\mathcal{M}^t$ represents a pair $\ell_i,\ell_j$, then the $v^{th}$ column entry of the corresponding row in $\mathcal{M}^t$ is an \textit{exclusive OR} of the $(i,v)^{th}$ and $(j,v)^{th}$ entries of the influence matrix $\mathcal{M}$.
The above point illustrates the fact that to localize an event $\ell_i$, there always exists a sensor that distinguishes $\ell_i$ from $\ell_j$ by producing different outputs for $\ell_i$ and $\ell_j$ respectively, i.e., if a sensor output is $1$ (resp. $0$) in case of $\ell_i$, then its output for $\ell_j$ is $0$ (resp. $1$), for all $j\ne i$.  \vspace{-0.25cm}

Note that for a given subset of sensors $S$, the identification function, which is the number of pair-wise link failures detected by $S$, is essentially same as the detection function of $S$ in the corresponding MSC instance i.e., \vspace{-0.35cm}

\begin{equation} \label{eq:objidnt}
f_I(\mathcal{C}_S) =f_D(\mathcal{C}_S^t),
\end{equation}\vspace{-0.35cm}


where $f_D$ is defined as in \eqref{eq:6}.
The \textit{normalized identification score}, denoted by $I_I(S)$, is computed by dividing $f_I$ by the total number of pair-wise events, $|\mathcal{L}^t|$.

\subsubsection{Greedy approach based solution} \vspace{-0.35cm}
Once the MTC problem has been transformed to the MSC problem, a straightforward way to obtain a solution is to employ the greedy algorithm, as outlined in Algorithm \ref{alg:1}.

\begin{algorithm} 
\small
\caption{Minimum Test Cover -- Greedy Algorithm}\label{alg:1}
\begin{algorithmic}[1]
\State \textbf{Input:} $\mathcal{C} = \{C_1,\cdots,C_m\},\;C_i\subseteq\mathcal{L}$
\State \textbf{Output:} MTC: $\mathcal{C}^{{\ast}}\subseteq\mathcal{C}$
\State \textbf{Initialize:} $\mathcal{C}^{{\ast}}\gets \emptyset$
\State \textbf{Transform:} the test cover instance to the set cover instance, i.e., from a given $\mathcal{L}$ and $\mathcal{C}$, obtain a corresponding $\mathcal{L}^t$ and $\mathcal{C}^t$ (Section \ref{sec:t2s}).
\State \textbf{Solve:} using greedy algorithm\label{algo:greedy}
	\begin{enumerate}[(a)]
\item
Select $C_{i^\ast}^t\in\mathcal{C}^t$ (i.e., the sensor $i^\ast$) covering the most uncovered elements in $\mathcal{L}^t$. 
\item Add to current set $\mathcal{C}^{{\ast}}\gets \mathcal{C}^{\ast} \cup \{C_{i^\ast}\}$.
\item Repeat until all elements in $\mathcal{L}^t$ are covered, or no new element in $\mathcal{L}^t$ can be covered by any $C_i^t\in\mathcal{C}^t$.
\end{enumerate}
\end{algorithmic}
\end{algorithm}

As in the case of the MSC problem, the lazy greedy approach, which exploits the submodularity property of the set cover problem, can be utilized. However, if there are $n$ link failures that need to be localized, then the corresponding set cover instance contains $\dbinom{n}{2}$ events, and the time complexity of the greedy approach in Algorithm \ref{alg:1} is $\mathcal{O}\left(m\dbinom{n}{2}\right)$, where $m$ is the total number of sensors. Even for small-sized networks with a limited number of possible link failures, this approach becomes quite inefficient owing to a large number of computations required. Moreover, employing lazy greedy also achieves desired computational efficiency for realistic size of failure event set. In the next section, we focus on improving the computational time of the solution of the MTC problem.

\section{Augmented greedy MTC solution}
\label{sec:New_Algo}\vspace{-0.35cm}
The main idea behind the augmented greedy approach is to achieve a computationally efficient approximation algorithm. We do so by avoiding the complete transformation of the MTC to the MSC and directly evaluating the function \eqref{eq:objidnt}, thus eliminating the need to \textit{pre}-compute the identification matrix $\mathcal{M}^t(\mathcal{L}^t,\mathcal{S})$. For example, for a network with $m = 2000; n = 2000;$ we would require $\sim 4 \text{ } GB$ computer memory to store the transformed MSC. \vspace{-0.25cm}

In each iteration of the greedy algorithm for the MTC solution, a sensor that covers (detects) the most pair-wise link failures from a total of $\dbinom{n}{2}$ pair-wise failures, is selected. Thus $\mathcal{O}\left(\dbinom{n}{2}\right)$ comparisons are made in a single iteration for each potential sensor. In the augmented greedy approach, we avoid this by significantly reducing the number of comparisons made in each step. In fact, for each sensor, the number of comparisons made in a single iteration are always bounded by $\mathcal{O}\left(K\dbinom{k}{2}\right)$, where $k$ is the maximum number of link failures that are detected by any sensor, and $K$ is the number of sensors that are included in the test cover until that iteration. Since $k$ is typically much smaller than $n$, a large number of computations are thus avoided in each iteration.  \vspace{-0.25cm}

\begin{figure}[ht]
\centering
\includegraphics[scale = 0.35]{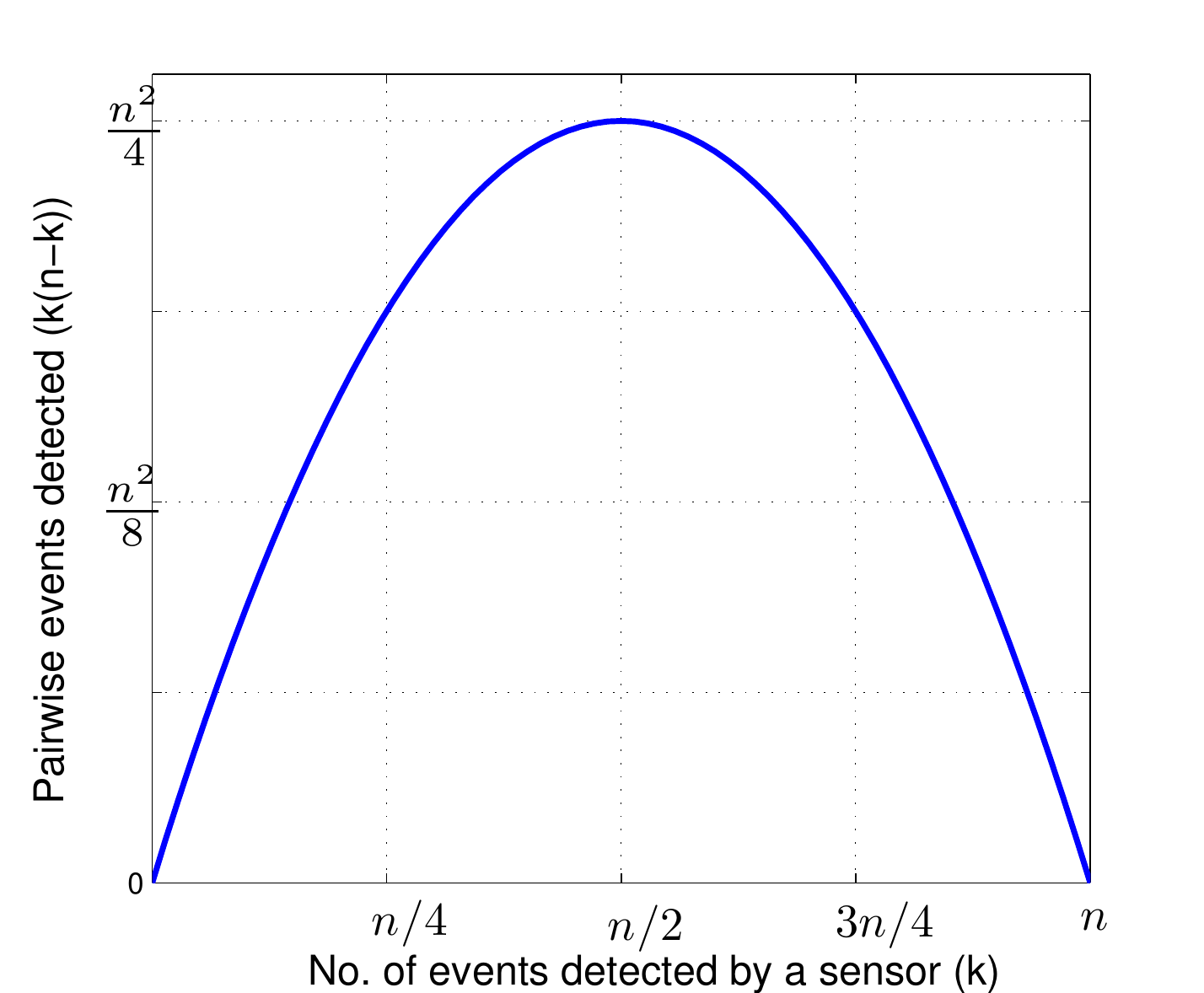}
\caption{The number of pair-wise link detections as a function of the number of detected events.}
\label{fig:knk}
\end{figure}

To explain our approach, we first observe that a sensor $i$ that detects $k$ events (i.e., $\lvert C_i \rvert = k$) can distinguish between $k$ detected events and $(n-k)$ undetected events. Thus, such a sensor detects $k(n-k)$ \textit{pair-wise} events (i.e., $\lvert C^t_i \rvert = k(n-k)$). Unlike the detection problem, in which a sensor with a large $k$ is desirable for the detection purposes, a sensor that detects a large number of failures is not always useful for the identification.
Figure \ref{fig:knk} shows the number of pair-wise events detected by a sensor as a function of the number of (single) events detected by the sensor. The maximum number of pair-wise events, which are link failures in our case, are detected when $k=n/2$. 

Moreover, if a sensor $i$ included in a test cover and $\ell_u,\ell_v \in C_i$, then a distinction between the occurrence of $\ell_u$ and $\ell_v$ is not possible through the sensor $i$.
Thus, a set of sensors that can distinguish between events $\ell_u,\ell_v\in C_i$, or equivalently that can detect pair-wise events corresponding to the events in $C_i$, also need to be included in the test cover. Based on this observation, we suggest an augmented greedy approach to compute the test cover without computing the $\dbinom{n}{2}$ events priori. \vspace{-0.25cm}

Let $C^{\ast}\subseteq\mathcal{C}$ be the test cover until the current iteration, and $\mathcal{C}_{cov}$ be the set of link failures detected by the sensors that are included in the test cover, i.e., $\mathcal{C}_{cov} = \bigcup\limits_{C_u\in C^\ast}C_u$. Thus, the utility of adding $C_i$ to $C^\ast$ (i.e., adding sensor $S_i$ to the test cover) in each iteration is based on the following two factors: \vspace{-0.15cm}

\begin{enumerate}[(i)]
\item
How many pair-wise link failures corresponding to the links which are \textbf{not included} in $\mathcal{C}_{cov}$ can be detected by $C_i$? We define this value as $x_i$.
\item
How many pair-wise link failures corresponding to the links already \textbf{included} in $\mathcal{C}_{cov}$ can be detected by $C_i$? We define this value as $y_i$.
\end{enumerate}

The overall utility of adding sensor $S_i$ to the test cover, denoted by $w_i$, is the sum of $x_i$ and $y_i$. A sensor $S_{i^\ast}$ that maximizes this overall utility, let $w_{i^\ast}$ denote the maximum utility, will then be included in the test cover, and $\mathcal{C}_{cov}$ will be updated to $\mathcal{C}_{cov}\gets \mathcal{C}_{cov}\cup C_{i^\ast}$. Now, we state how to compute $x_i$ and $y_i$ in the $j^{th}$ iteration. \vspace{-0.25cm}

\begin{enumerate}[(i)]
\item
\textit{Computing $x_i$} -- If $n_j$ is the number of link failures that are not yet included in $\mathcal{C}_{cov}$, (i.e., $n_j = n- \lvert\mathcal{C}_{cov}\rvert$), and $C_i$ contains $k_{i,j}$ of such link failures, then $x_i =~k_{i,j}(n_j-~k_{i,j})$. Note that computing $x_i$ is very straight forward and does not require computing pair-wise link failures from a given set of link failures.

\item \textit{Computing $y_i$} -- If a sensor $u$ is already included in the test cover, then the pair-wise link failures corresponding to the links in $C_u$ remain undetected. Thus, $y_i$ computes how many of such pair-wise link failures can be detected by the inclusion of sensor $i$ in the test cover. To make it precise, we proceed as follows: \vspace{-0.25cm}
\end{enumerate}

If $X$ and $Y$ are two sets, then we define: \vspace{-0.35cm}
$$\beta(X) =  \text{set of all 2-element subsets of}\; X,$$ \vspace{-0.35cm}
and \vspace{-0.35cm}
\begin{equation*}
\label{eq:beta}
\alpha(Y,\beta(X)) = \{a\in\beta(X):\;\lvert Y \cap a \rvert = 1\}.
\end{equation*} 
Here, $\alpha(Y,\beta(X))$ is a set consisting of such $2$-element subsets of $X$ that have exactly one common element with $Y$. For instance, if $X=\{1,2,3\}$ and $Y=\{1,3\}$, then $\beta(X)=\{\{1,2\},\{1,3\},\{2,3\}\}$, and $\alpha(Y,\beta(X))=\{\{1,2\},\{2,3\}\}$. \vspace{-0.25cm}

To compute $y_i$, first we compute the set of link failures common to $C_i$ and $\mathcal{C}_{cov}$ and call it as $Y_i=C_i\cap\mathcal{C}_{cov}$. Now, if sensor $u$ is already included in the test cover, and $G_u\subseteq\beta(X_u)$ is the set of undetected pair-wise link failures corresponding to the links in $X_u\subseteq C_u$, then  \vspace{-0.35cm}
$$
y_i = \sum\limits_{C_u\in C^{\ast}}\left\vert \alpha(Y_i, G_{u})\right\vert \vspace{-0.35cm}
$$ 
The complete algorithm is stated in Algorithm \ref{euclid}.

\begin{algorithm}
\small
\label{algo:fast}
\caption{Minimum Test Cover -- Augmented Greedy Algorithm}\label{euclid}
\begin{algorithmic}[1]
\State \textbf{Input:} $\mathcal{C} = \{C_1,\cdots,C_m\},\;C_i\subseteq\mathcal{L}$
\State \textbf{Output:} MTC: $\mathcal{C}^{{\ast}}\subseteq\mathcal{C}$
\State \textbf{Initialization:} $\mathcal{C}_{cov} = \emptyset$;\; $\mathcal{C}^{{\ast}}=\emptyset $;\; $G_0=\emptyset$;\; $j=1$;\; $n = |\mathcal{L}|$; \; $w_{i^\ast}=1$;
\While {$w_{i^\ast} > 0$}
\State $n_j \gets n-\left\vert \mathcal{C}_{cov}\right\vert$
\For {all $i$}
\State  $X_i \gets \left(C_i \setminus \mathcal{C}_{cov}\right); k_{i,j}\gets \left\vert X_i\right\vert$
\State $x_i\gets k_{i,j}(n_j-k_{i,j})$
\State $Y_i\gets C_i\cap \mathcal{C}_{cov}$
\State $y_i \gets \sum\limits_{t=0}^{j-1}\left\vert \alpha(Y_i, G_{t})\right\vert$
\State $w_i = x_i + y_i$
\EndFor \textbf{end for}
\State $w_{i^{\ast}}\gets \max w_i$
\algrule
\If {$w_{i^\ast} > 0$}
\State $\mathcal{C}^{{\ast}}\gets \mathcal{C}^{{\ast}} \cup \{C_{i^{\ast}}\}$
\State $\mathcal{C}_{cov} \gets \mathcal{C}_{cov} \cup C_{i^{\ast}}$
\State $G_j \gets \beta(X_{i^{\ast}})$
\For {$t=0$ to $j-1$}
\State $G_t \gets G_t \setminus \alpha(Y_{i^{\ast}},G_t)$ 
\EndFor \textbf{end for}
\State $j\gets j+1$
\EndIf \textbf{end if}
\EndWhile \textbf{end while}
\end{algorithmic}
\end{algorithm}

\begin{exmp}[Augmented greedy]
Consider the network shown in Figure \ref{fig:3}. Let $k_i$ be the number of failure events detected by the sensor $i$, i.e., $|C_i| = k_i$, where $C_i\subseteq\mathcal{S}$. In the first iteration ($j=1$) of the while loop, size of the event space is $n=10$, and $k_{i,j}=k_i$, $\forall i$.  
Then, the number of new pair-wise link failures detected by the sensor $i$ is given by $x_i = k_{i,j}(n - k_{i,j})$. Since there are no sensors in the test cover in the first iteration, $y_i=0$ for all the sensors. The maximum value of $w_{i}$ is attained for the sensors $1$ and $2$ with $w_1=w_2=x_1=x_2=5(10-5)=25$. We include sensor $1$ in the test cover, thus $\mathcal{C}^\ast = C_1$ after the first iteration of the while loop. The set of all undetected pair-wise events for sensor $1$, $G_1 =  \left\{\{1,2\},\{1,3\},\cdots,\{4,5\} \right\}$,  are then updated. Finally, we update the number of covered events as $\mathcal{C}_{cov}= \{1,2,3,4,5\}$. For the second iteration, i.e., $j=2$, size of the event space is updated as $n_2 = 5$.
A complete account of the states of variables of the algorithm for the example is provided in the \cite{perelman2015}.
The algorithm returns the test cover consisting of sensors $\{1,2,3,5\}$ that uniquely identify all link failures. 
\end{exmp}

The augmented greedy approach in Algorithm \ref{euclid} produces the same solution as the greedy approach in Algorithm \ref{alg:1}. Thus, Algorithm \ref{euclid} has the same approximation ratio as the standard greedy algorithm, which has been proven to be the best possible.  \vspace{-0.25cm}

Since a large number of computations are avoided in the execution of Algorithm \ref{euclid}, it is more efficient than the simple greedy. In contrast to the $\mathcal{O}\left(\dbinom{n}{2}\right)$ comparisons performed in each iteration for a sensor in Algorithm~\ref{alg:1}, $\mathcal{O}\left(\sum\limits_i^{m_j} \dbinom{k_i}{2}\right)$ comparisons are done in each iteration of the Algorithm~\ref{euclid}. Here, $n$ is the total number of link failures, $k_i$ is the number of link failures detected by the sensor $i$ (i.e., $k_i=|C_i|$), and $m_j$ is the number of sensors included in the test cover until that iteration. Thus, if $k=\max({k_i})$, then Algorithm \ref{euclid} is at least $n/k$ times faster than the simple greedy approach as shown below. Moreover, typically $k<<n$ in the case of link failure detection in water distribution networks, thus, $n/k$ factor turns out to be a significant improvement.
\begin{proposition}
\label{prop:compare}
Let $\sum\limits_i k_i= n$, and $k=\max (k_i)$, then 
\begin{equation}
\label{eq:one}
\sum\limits_{i}{\dbinom{k_i}{2}}\le \frac{k}{n}\dbinom{n}{2}
\end{equation}
\end{proposition}\vspace{-0.35cm}
\textit{Proof --} \vspace{-0.35cm}
\begin{equation*}
\begin{split}
\sum\limits_{i}{\dbinom{k_i}{2}}  & = \frac{1}{2}\left(\sum\limits_i k_i^2 - \sum\limits_i k_i\right) \le \frac{1}{2}\left(k\sum\limits_i k_i -n\right) \\
& = \frac{1}{2}\left(kn - n\right)\le \frac{1}{2}\left(kn-k\right) =\frac{k}{n}\dbinom{n}{2}.\qed
\end{split}
\end{equation*}

We note that Algorithm \ref{euclid} is somewhat similar to the two-step greedy algorithm presented in \cite{Bontriddr1}. However, in our approach, both $x_i$ and $y_i$ are computed in the same iteration resulting in a more efficient implementation.

\section{Application to a benchmark network}
\label{sec:applications}
\begin{figure}[ht]
\centering
\includegraphics[scale = 0.35]{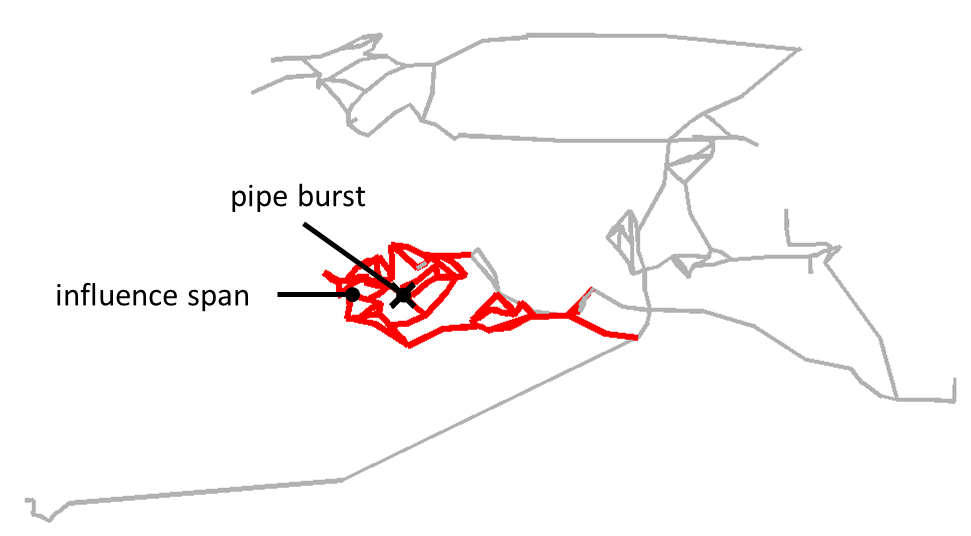}
\caption{Layout of Net1 and propagation of failure in LINK-126}
\label{fig:net1}
\end{figure}

\vspace{-0.35cm}
We first test our approach on a medium-size water network. \textit{Net1} is a benchmark system that has been extensively studied in the context of sensor placement for water quality  \cite{ostfeld}. The system consists of 126 nodes, 168 pipes, one reservoir, one pump, and two storage tanks and its layout is shown in Figure \ref{fig:net1}. The system supplies a daily demand of $5.15\times10^3 [m^3/day]$ and has a total pipe length of $37.5\times10^3[m]$.  \vspace{-0.35cm}

For all our simulations, we consider a single failure event occurring at the center of each pipe and enumerate all possible failure events. For the detection problem, when fully calibrated transient model of the network is not available, we approximate the disturbance propagation using a simple distance based model emulating the dissipation of the pressure wave with the distance from the origin. As in \cite{fd:deshpande}, our influence model is based on the shortest distance threshold model, assuming that the disturbance in pressure can be sensed within a specified distance from the location of the burst, i.e., $\mathbf{y}_{S_i}(\ell_j)= \{ 1 \hspace{0.5em} | \hspace{0.5em} d (S_i,\ell_j) \leq \varepsilon \}$, where $d$ is the length of the shortest path between two locations $S_i$ and $\ell_j$, and $\varepsilon$ is some threshold.  
Figure \ref{fig:net1} shows an example of the influence range (in red) of a burst in LINK-126 of the network for a threshold distance of $\varepsilon = 1000 [m]$, i.e., a sensor located in the red region can detect the pipe failure.

\begin{figure} 
	\centering
		\includegraphics[scale = 0.2]{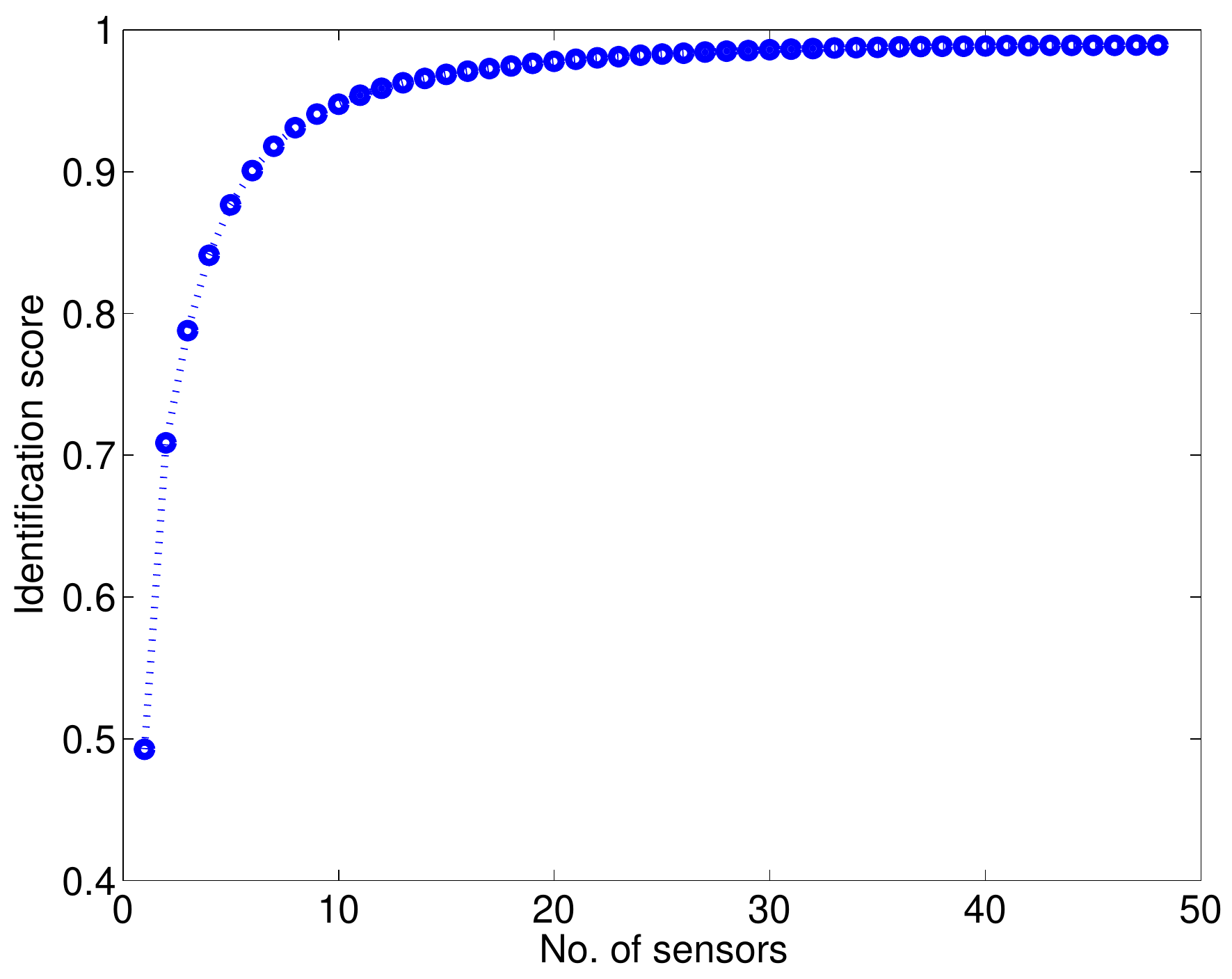}       
		\caption{\footnotesize Identification score for Net1}   \label{fig:net1_1} 
	\end{figure}

Assuming that a sensor can be placed at any of the 126 network nodes and any of the 168 network pipes can fail, we solve the MTC problem, as described previously in sections \ref{sec:Ident_as_MTC}, \ref{sec:3.2}, and \ref{sec:New_Algo}.  
Figure \ref{fig:net1_1} shows the normalized identification score, $I_I$, defined in Section \ref{sec:t2s}, as a function of the number of sensors using the greedy approach. 
As noted in Section \ref{sec:ds}, we observe that the identification score function exhibits a diminishing return property. The maximum identification score of 0.99 is attained with 48 sensors.

Observing that the identification score of the network is not sufficient to evaluate the quality of the design, since it does not indicate about the number of events that are uniquely identified and, respectively, the number of events that are not uniquely identified. For this reason, we suggest two complementary metrics for evaluating the performance of the sensor network design: \vspace{-0.25cm}

\emph{Localization score} -- Let $L\subseteq\mathcal{L}$ be a subset of all such link failures for which the outputs of sensors in $S$ is same, i.e., $ \mathbf{y}_S(\ell_i) = \mathbf{y}_S(\ell_j)$, $\forall \ell_i=\ell_j \in L$. We call such a subset of link failures $L$ as a \textit{localization set}. A localization can be associated with every unique vector of sensors' outputs. Localization score is the total umber of localization sets obtained under the sensor configuration $S$.  
We note that it is not possible to distinguish between the failure events in a localization set by merely observing the outputs of sensors. We define the normalized localization score, $I_L(S)$, as the ratio of the total number of localization sets formed under the sensor configuration $S$ to the total number of event failures. 
Ideally, the normalized localization score should be equal to 1, indicating that each fault can be uniquely identified. 
	
\begin{figure}
        \centering
        \begin{subfigure}[b]{0.23\textwidth}
               \includegraphics[scale = 0.2]{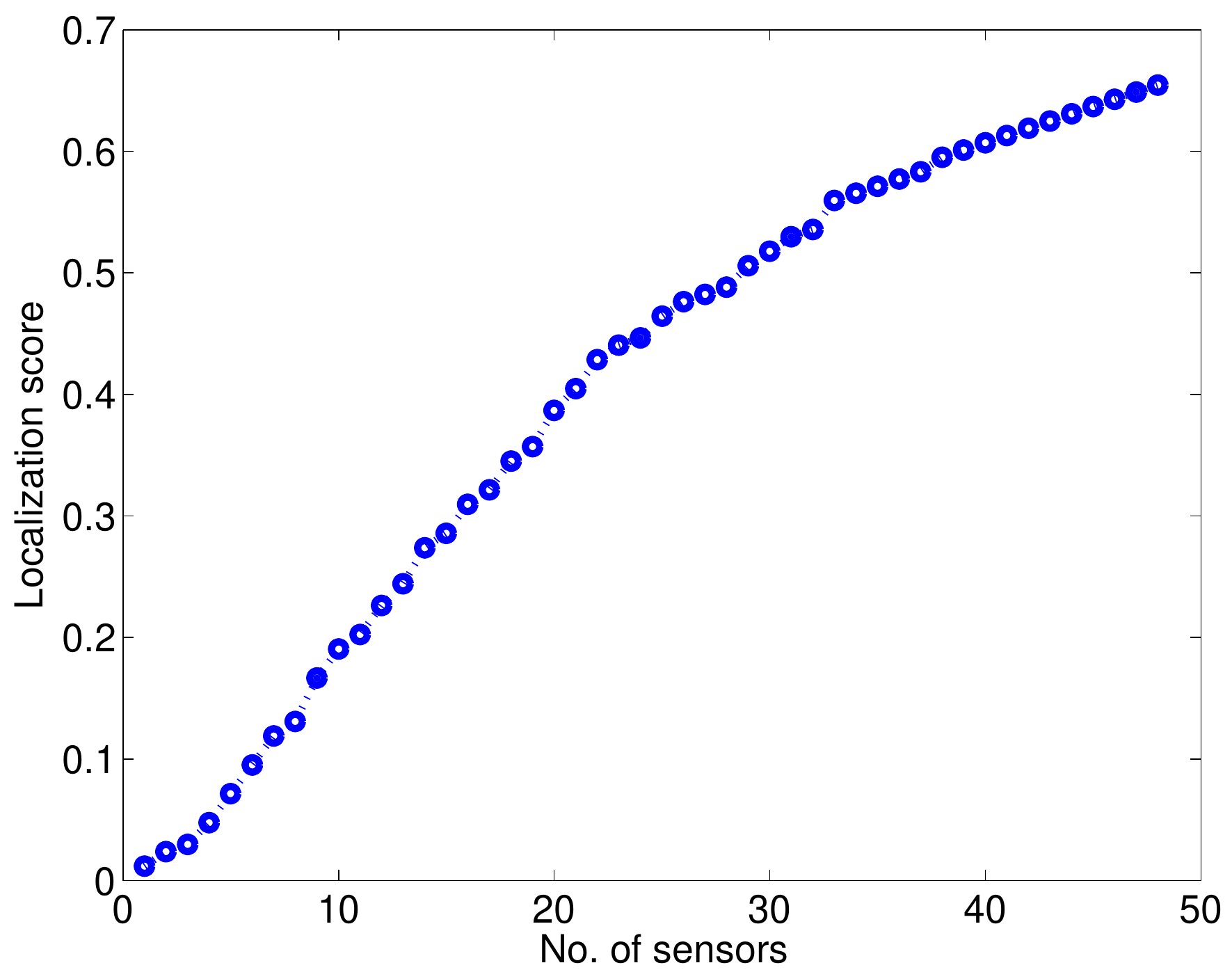}       
		
		\caption{\footnotesize Localization score}    \label{fig:net1_2} 
        \end{subfigure}
        ~ 
        \begin{subfigure}[b]{0.23\textwidth}
                \includegraphics[scale = 0.2]{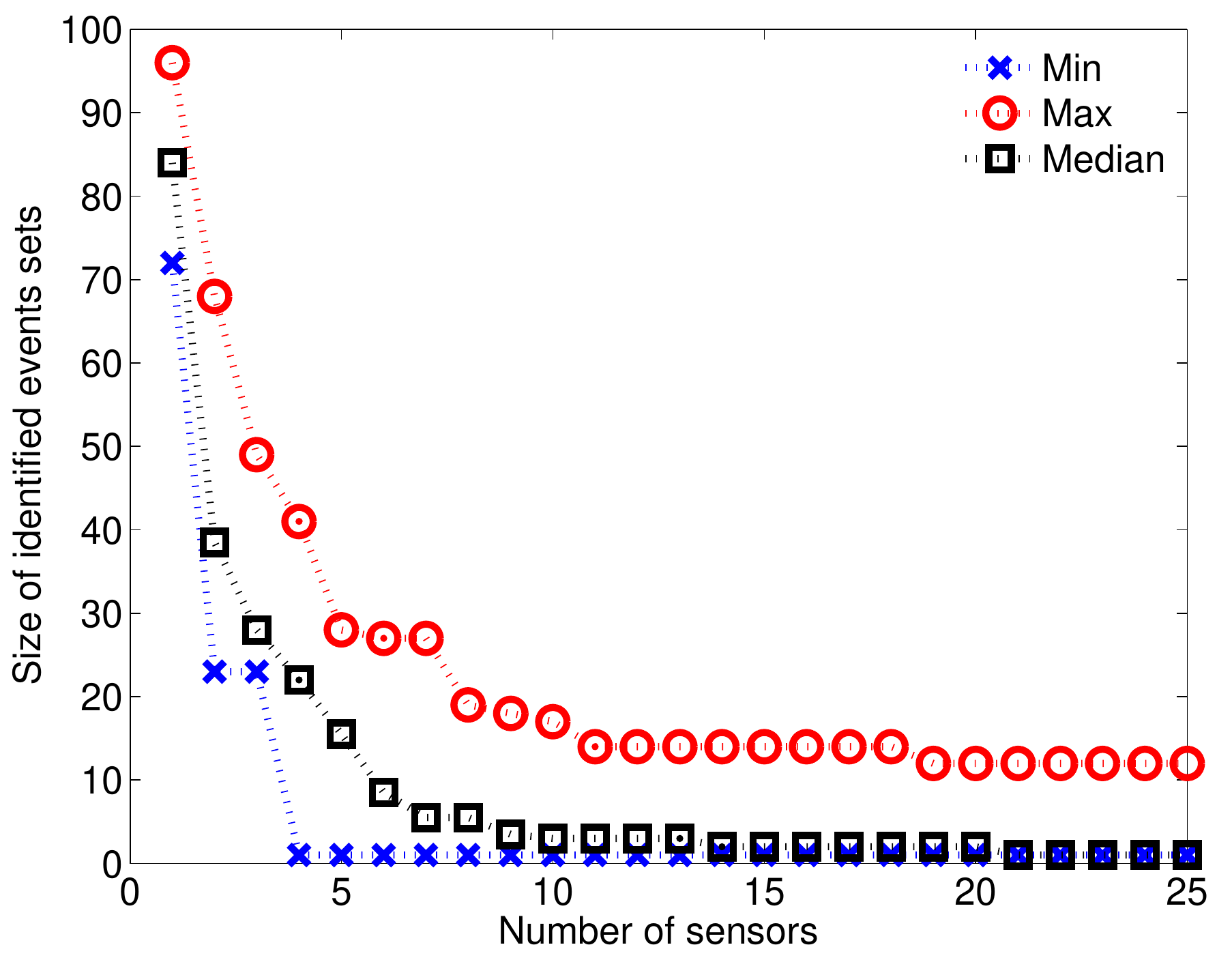}        
		\caption{\footnotesize Localization-set-size}   \label{fig:net1_3}  
        \end{subfigure}
       \caption{Localization performance for Net1}  \label{fig:net2_2}
\end{figure}
\vspace{-0.25cm}
\emph{Localization size} -- is the number of faults associated with a unique output of sensors, or the number of elements in a localization set $L$. A localization size of higher value means  that it would be difficult to identify the location of the fault, and additional local inspection methods might be needed. We define the worst set size, $I_W(S)$ as the largest localization set. For complete localization it is required that, $I_W(S)=1$, indicating that all faults could be distinguished from each other, and therefore could be uniquely detected. \vspace{-0.35cm}

\begin{exmp}[Localization score]
Continuing Example \ref{exmp:2} for the two-sensor design $ S= \{S_2,S_4 \}$, three localization sets are formed, i.e. $L_1 =\{ \ell_1 \}, L_2 = \{ \ell_4,\ell_5,\ell_7, \ell_9,\ell_{10}\}, L_3  = \{ \ell_2,\ell_3, \ell_6,\ell_8 \}$. The corresponding localization sizes are $|L_1| = 1, |L_2| = 5, |L_3| = 4$. The normalized localization score is thus $I_L = 3/10$ and the worst localization size is $I_W = 5$. It means that if an event is detected, its distinction between three distinct groups is possible, but further distinction within the groups is not possible, with the largest indistinctive group of 5 links. 
With the four-sensor design, $ S^*= \{S_1,S_2,S_3,S_5 \}$, the optimal normalized localization score and the maximum localization size of 1 are achieved, and we observe ten unique outputs of sensors, each associated with a unique failure event.
\end{exmp}\vspace{-0.25cm}

Figure \ref{fig:net1_2} shows the normalized localization score as a function of the number of sensors. The highest localization score of 0.65 is achieved when 48 sensors are installed. This result indicates that 110 unique vectors of sensors output are associated with the 168 failure events. 
Figure \ref{fig:net1_3} shows the worst, median, and minimum localization set sizes as a function of the number of sensors for Net1. We observe that initially sizes of localization sets decrease rapidly with the number of sensors, until the worst localization-set-size reaches a plateau at 20 sensors, and does not improve further. This implies that deploying more sensors might improve local performance, but will not improve the overall network localization performance, making further deployment of sensor unattractive for the water utility from the cost viewpoint. 

\section{Application to real networks} \label{sec:app2}\vspace{-0.35cm}

We tested our approach on a batch of real water networks. Principal information is listed in Table \ref{tab:1} and the complete data can be obtained from \cite{doi:10.1061/(ASCE)WR.1943-5452.0000352} for Nets 2-10 and from \cite{exeter} for Nets 1,11,12. In all our simulations we again assume, that a single failure can occur at each of the network links and that sensors can be placed at each of the network nodes, and set the distance threshold to $\varepsilon = 1000 [m]$. 

\begin{table}[htbp]
  \centering

  \caption{Network data}
\tabcolsep=0.11cm
	\renewcommand{\arraystretch}{1.1} 
    \begin{tabular}{c|cccc}
   	 \multirow{ 2}{*} {\textbf{Network}} &  \textbf{Length} &  \textbf{Demand} & \textbf{No. of}  &  \textbf{No. of} \\
    &  \textbf{$[km]$} & $10^3[m^3/day]$ & \textbf{pipes}   & \textbf{nodes} \\
\hline
   \textbf{Net1}   & 37.56 & 5.15& 168   & 126 \\
    \textbf{Net2}  & 91.29 & 7.59& 366   & 269 \\
    \textbf{Net3}  & 96.58 & 8.58& 496   & 420 \\
    \textbf{Net4}  & 137.05 &5.78&  603   & 481 \\
    \textbf{Net5}  & 123.20 & 6.20&644   & 543 \\
    \textbf{Net6}  & 166.60 & 5.66& 907   & 791 \\
    \textbf{Net7}  & 153.30 & 8.93& 940   & 778 \\
    \textbf{Net8}  & 152.25 & 7.91& 1124  & 811 \\
    \textbf{Net9}  & 260.24 & 5.67& 1156  & 959 \\
    \textbf{Net10} & 247.34 & 9.33& 1614  & 1325 \\
 \textbf{Net11} & 760.89 & 71.88& 3032 & 1891 \\
 \textbf{Net12}& 1844.04 & 108.8& 14822  & 12523 \\

    \end{tabular}%
  \label{tab:1}%
\end{table}%

\subsection{MSC vs. MTC} 
\begin{figure}
	\centering
		\includegraphics[scale = 0.5]{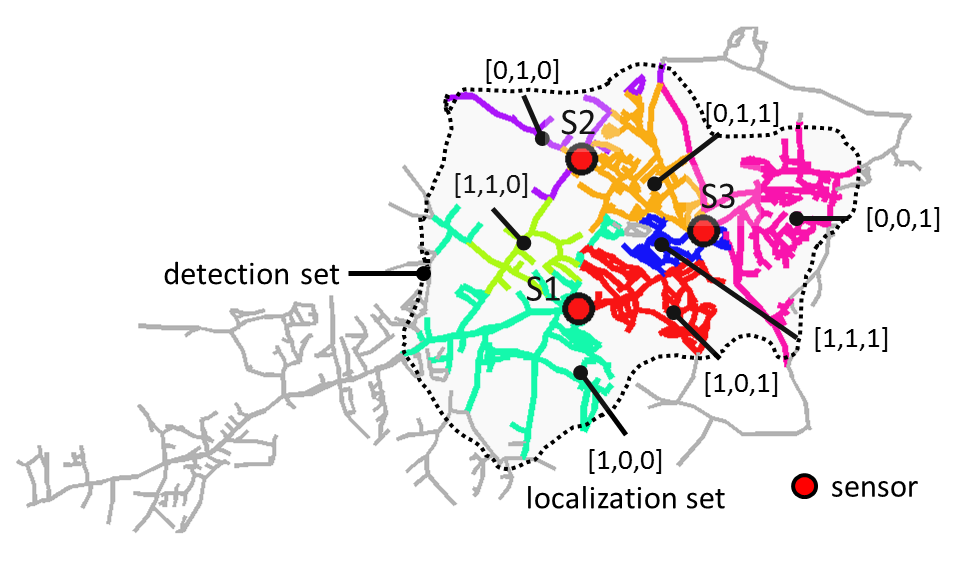}  
		\caption{\footnotesize Layout of Net9 and example of the detection and localization sets for three sensors}   		\label{fig:net2}
	\end{figure}

\vspace{-0.35cm}
First, we compare the sensor placement design for the identification problem obtained from our approach with the design for the detection problem, i.e. \textit{MTC vs. MSC} (Sections \ref{sec:dmsc}, \ref{sec:Ident_as_MTC}). We demonstrate our results using \emph{Net9}, from the Kentucky dataset. Although the system supplies similar daily demand as Net1, it is spatially more distributed with approximately 260 $[km]$ of pipes. Network layout and main features are shown in Figure \ref{fig:net2} and Table \ref{tab:1}. \vspace{-0.35cm}

Figure \ref{fig:net2} schematically illustrates the difference between the MTC and MSC problem formulations in the context of Net9. Consider three sensors installed in the network, Figure \ref{fig:net2} demonstrates the seven localization sets corresponding to seven unique sensor states, $[0,0,1],\cdots, [1,1,1]$ and the detection set, being the union of the localization sets. Whereas the detection problem tries to maximize the detection set, the identification problem aims to identify distinct subsets.

\vspace{-0.15cm}

\begin{figure}
        \centering
        \begin{subfigure}[b]{0.23\textwidth}
                \includegraphics[scale=0.23]{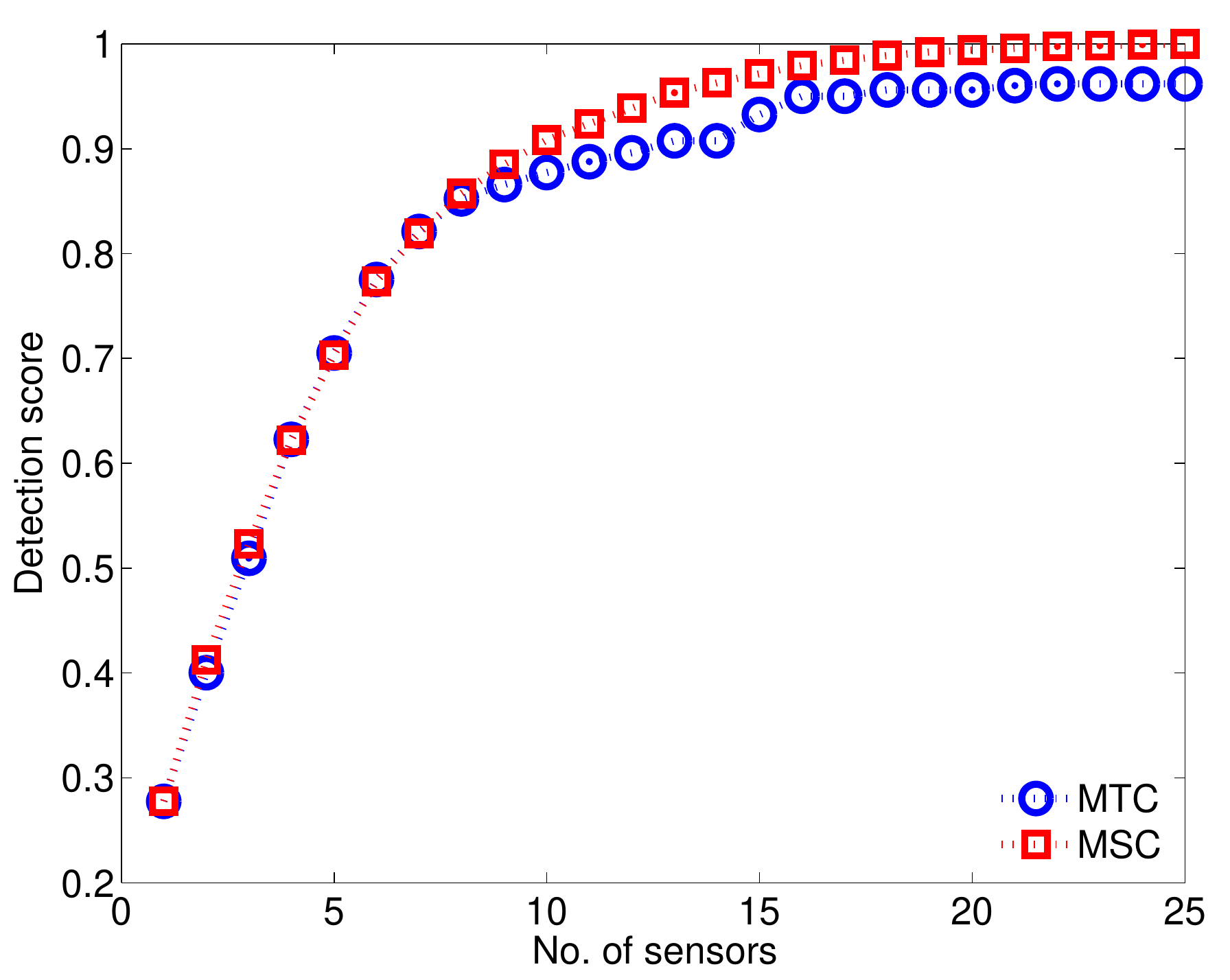}
		\caption{Detection}
                \label{fig:net2_2a}
        \end{subfigure}%
        ~ 
        \begin{subfigure}[b]{0.23\textwidth}
                \includegraphics[scale=0.23]{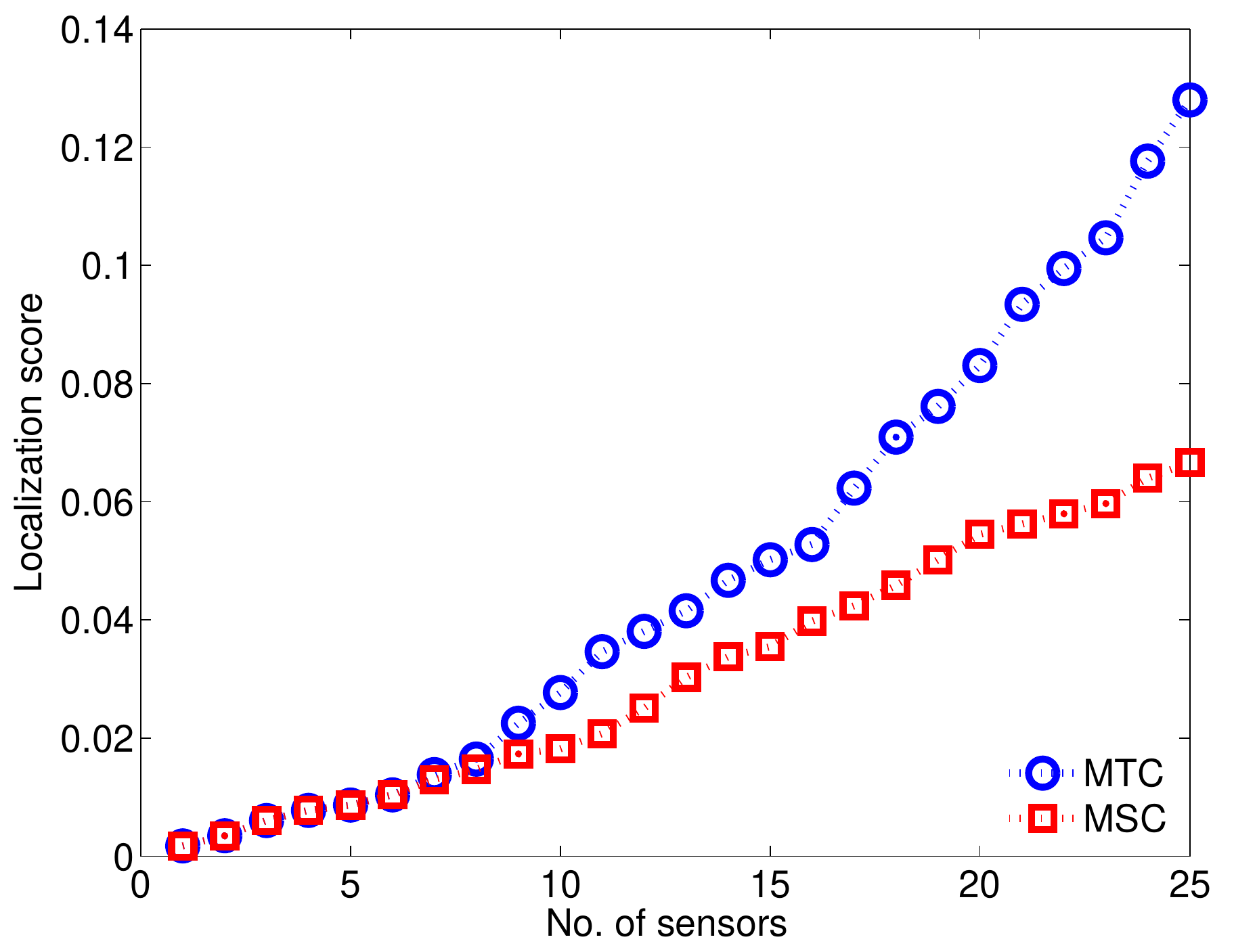}
		\caption{Localization}
                \label{fig:net2_2b}
        \end{subfigure}
       \caption{MTC versus MSC performance for Net9}  \label{fig:net2_2}
\end{figure}

Figure \ref{fig:net2_2} provides a comparison between the detection and localization scores for the MTC (blue circles) and MSC (red squares) designs. 
For the detection problem, 25 sensors are sufficient to cover the entire system, hence, we also select the first 25 sensors for the identification problem and compare their performance. From Figure \ref{fig:net2_2a} it can be seen that the two designs overlap for the first 7 sensors and the MSC design only slightly outperforms the MTC design when comparing the detection scores for a higher number of sensors. At the same time, the MTC design significantly outperforms the MSC design when comparing the localization scores as shown in Figure \ref{fig:net2_2b}. Similar results were attained for the other networks.

\subsection{Augmented greedy vs. transformed lazy greedy}\vspace{-0.35cm} \label{sec:app}
Next, we compare the solution approach based on the augmented greedy (AG) (Section \ref{sec:New_Algo}) and the transformed lazy greedy (TLG) (Section \ref{sec:3.2}). Table \ref{tab:2} lists the running times (Intel Core i7, 2.9 GHz, 16 GB of RAM) for the augmented greedy and the transformed lazy greedy approaches. For Nets 1-10, the new algorithm is $3$ to $8$ times faster than the transformed lazy greedy approach, depending on the maximum number of events detected by any sensor (see Proposition \ref{prop:compare}). The solutions obtained using the two approaches were identical. {For Nets 11-12, we were not able to apply the TLG due to the memory requirements and applied only the AG, which further emphasizes the advantage of the AG approach.}\vspace{-0.15cm}

Finally, Table \ref{tab:2} lists the maximum number of sensors and the corresponding four performance scores: normalized detection $I_D$, identification $I_I$, and localization $I_L$ scores, and worst localization set size $I_W$. For all networks, the layouts and the simulation plots illustrating these metrics as a function of the number of sensors are available in \cite{perelman2015}. These results demonstrate that: (1) The number of sensors required solely for detection purpose is significantly lower than the number of sensors required for localization. 
(2) Between the two localization measures, $I_L$ and $I_W$, the localization score is more conservative than the worst set size, requiring a larger number of sensors.  
For example, consider the design for Net9, then to detect 95\% of the events, i.e., $I_D = 0.95$, 18 sensors are sufficient, whereas to achieve $I_L = 0.5$ we require 79 sensors, and 38 to achieve $I_W = 20$. This is observed for all tested networks. 

\begin{table}[htbp] 
  \centering
  \caption{Simulation results}
\tabcolsep=0.13cm
	\renewcommand{\arraystretch}{1.1} 
    \begin{tabular}{c|c|cccc|cc}
	 \multirow{ 2}{*} {\textbf{Network}} & \textbf{No. of} &  \multirow{ 2}{*} {$I_D$}  & \multirow{ 2}{*} {$I_I$}  & \multirow{ 2}{*} {$I_L$}& \multirow{ 2}{*} {$I_W$} &\textbf{TLG} & \textbf{AG} \\
    &  \textbf{sensors} &  & & &  & \textbf{$[min]$} & \textbf{$[min]$} \\
\hline
    \textbf{Net1} & 48    & 0.99  & 0.99  & 0.65  & 12    & 0.23  & 0.08 \\
    \textbf{Net2}  & 98    & 0.99  & 1.00  & 0.86  & 12    & 2.39  & 0.58 \\
    \textbf{Net3} & 134   & 0.99  & 1.00  & 0.86  & 7     & 6.93  & 1.65 \\
    \textbf{Net4}  & 138   & 0.99  & 1.00  & 0.91  & 8     & 11.98 & 4.93 \\
    \textbf{Net5}  & 164   & 0.99  & 1.00  & 0.86  & 6     & 15.58 & 3.85 \\
    \textbf{Net6}  & 258   & 1.00  & 1.00  & 0.86  & 8     & 45.46 & 6.31 \\
    \textbf{Net7} & 139   & 1.00  & 1.00  & 0.83  & 8     & 49.12 & 9.31 \\
    \textbf{Net8}  & 195   & 1.00  & 1.00  & 0.70  & 8     & 80.55 & 28.07 \\
    \textbf{Net9}  & 359   & 1.00  & 1.00  & 0.87  & 6     & 91.57 & 11.06 \\
    \textbf{Net10} & 408   & 1.00  & 1.00  & 0.89  & 14    & 257.41 & 39.48 \\
{\textbf{Net11}} & 717   & 1.00  & 1.00  & 0.69  & 9    & -- & 50.53 \\
{\textbf{Net12}}& $1000^*$   & 1.00  & 1.00  & 0.38  & 17    & -- & 1800  \\
    \end{tabular}\vspace{0.1cm}
TLG - transformed lazy greedy; AG - augmented greedy;\\$^*$terminated after 1000 iterations%
  \label{tab:2}%
\end{table}%


\section{Related work}\label{sec:review}\vspace{-0.35cm}

\textit{Event detection in water networks.}
In the urban water sector, majority of previous works focused on the sensor placement for detecting hypothetical contamination events assuming perfect sensors capable of detecting all types of contaminants \cite{berry,5350461}. In a related work \cite{Krause08efficientsensor}, to detect the presence of contaminants in large water distribution systems, the notion of penalty reduction function was introduced to realize various objective functions such as reduction of detection time and the expected population affected. Submodularity of the penalty reduction function was then used to solve sensor placement problems efficiently and with provable guarantees. Moreover, various data and model-driven techniques also exist that are applied for system's state estimation and event detection and isolation \cite{eliades2014contamination,rosich2012fault}. The basic premise in these methods is that once the sensors are in place, data is collected and transmitted in real-time. The difference between measurements, such as pressure \cite{perez2014leak} and flow \cite{ragot2006fault}, and their estimated values obtained using the network hydraulic model, is then computed. 
Model based leakage detection techniques are employed primarily on the operational side with the objective to efficiently utilize available measurements along with the available system model to determine the system faults. \vspace{-0.35cm}

Our approach is somewhat related to \cite{fd:deshpande,fdi:sarrate}, which consider pipe bursts as failure events. In \cite{fd:deshpande}, detection of events in networks is studied using distance decaying sensing function. The problem is formulated as a continuous $p$-median facility location problem and solved using a gradient descent algorithm. However, in contrast to \cite{fd:deshpande}, in which only the detection problem is considered, we consider detection as well as location identification of link failures.  In \cite{fdi:sarrate} both the detection and location identification of failure events are considered in the problem formulation. \vspace{-0.35cm}

In this work, we consider the placement of online high-rate pressure sensors. Additional surface and inline detection techniques include acoustic, umbilical, and autonomous robots. These tools are principally used to verify and pinpoint the location of the burst, their operation is typically time consuming and expensive, and they are not suitable for continuous operation \cite{6901220}. Ideally, flow meters can also be used for detecting and localizing leaks in water networks. However, these are more expensive and can be typically installed on main pipelines only at the inlets of sub-networks \cite{Narayanan:2014:OMF:2602339.2602346}. Furthermore most flow meters do not react instantaneously to changes in flow, hence are more suitable for persistent leaks \cite{doi:10.1080/15730621003610878}.

\textit{Approximation algorithms.} The sensor placement problem is not unique to the water sector and can be found in many engineering applications. Sensor placement is in essence a combinatorial optimization problem, in which a minimum number of sensors are deployed to minimize the uncertainty about the events of interest. 
The dominant approach is to cast the sensor placement problem as the classical \textit{minimum set cover} (MSC) problem, in which given a set of $n$ elements and a collection of $m$ subsets, the goal is to select as few subsets as possible such that their union covers all elements. 
The MSC problem is known to be NP-hard \cite{Moret1}. The \textit{greedy algorithm} guarantees the best possible approximation ratio of $(\ln n + 1)$.
A key feature in the efficient and practically feasible greedy algorithm is exploiting the submodular property, i.e. decreasing marginal utility of the objective function. Extensive literature exists on the greedy approximation for submodular functions.
In \cite{Krause:2008:NSP:1390681.1390689}, a mutual information criterion was proposed to select the most informative sensors to monitor a spatial phenomenon modeled by a Gaussian process. The submodularity property of the criterion, as shown in \cite{nemhauser1978analysis}, was then exploited to obtain a polynomial time algorithm guaranteeing a constant factor approximation of the optimal sensor set. \vspace{-0.15cm} 

\textit{Model-based diagnosis.} Fault detection and identification (FDI) and consistency based diagnosis (DX) are two distinct approaches which rely on computing sets of events in a faulty system based on the discrepancies between the observed and predicted system behavior \cite{1335513}. In the FDI community fault diagnosis is captured by localizing faults based on residuals that capture these faults. The problem is then to select a set of residual generators that are sensitive to the set of faults \cite{krysander2008sensor,Raghuraj,svard2013realizability}. In the DX community, the diagnosis is derived by computing a set of conflicts that capture the faulty components that explain the observed failures \cite{abreu2009low,de2011hitting,feldman2008computing}. To compute the minimum set of residual generators or the minimum set of conflicts, the problem often relies on the MSC or the minimum hitting set (MHS) formulation. The MSC problem is equivalent to the MHS, in which given the same input as in the MSC, the goal is to find the smallest subset of elements that \textit{hits} (i.e. has a non empty intersection) every subset \cite{1335513}.\vspace{-0.35cm}

{In previous works \cite{krysander2008sensor,Raghuraj,svard2013realizability} the isolation solution is obtained by first computing the set of all \textit{pair-wise} faults from a given set of faults, and then using greedy heuristics to solve the MSC or the MHS problems. This is similar to the TLG approach described in Section \ref{sec:3.2}. Computing all pair-wise events is the main computational bottleneck, especially when applied to large scale networks. The AG presented in Section \ref{sec:New_Algo} is a faster implementation of the greedy approach for the solution of the MTC. Its main feature is avoiding the transformation of the MTC to the MSC/MHS, which makes it more suitable for large-scale distributed systems, as demonstrated for Nets 11-12 in Table \ref{tab:2}.}

\section{Conclusions and future work}
\label{sec:conclusions} \vspace{-0.35cm}
In this work, we focused on the sensor placement for fault location identification in water networks. We cast the problem as the minimum test cover problem and suggested a fast solution approach. Additionally, we tested and analyzed the solutions using multiple performance criteria for a suite of real water networks. The outcomes of our approach could provide a better diagnosis of failure events in terms of improved localization and response to failure events in operational mode, and could significantly reduce potential physical losses and service disruptions in water networks. In this work we assumed perfect sensing information, future extension will include sensor placement robust to erroneous and corrupt data.
\vspace{-0.35cm}

\printnomenclature[0.4in]
\nomenclature[1h]{$h$}{hydraulic head}%
\nomenclature[1f]{$f_D$}{detection function}%
\nomenclature[1f]{$f_I$}{identification function}%
\nomenclature[1l]{$\ell_j$}{$j^{th}$ (failure) event}%
\nomenclature[1l]{$\ell_{ij}^t$}{unordered pair of (failure) events $\ell_i$ and $\ell_j$}%
\nomenclature[1l]{$\mathcal{L}$}{set of all (failure) events}%
\nomenclature[1l]{$\mathcal{L}^t$}{set of all pair-wise (failure) events}%
\nomenclature[1p]{$p$}{pressure}%
\nomenclature[1q]{$q$}{flow}%
\nomenclature[1s]{$S_i$}{the location of the $i^{th}$ sensor}%
\nomenclature[1s]{$\mathcal{S}$}{set of all sensors}%
\nomenclature[1ya]{$\mathbf{y}_{{\mathcal{S}}}$}{outputs of sensors in the set $\mathcal{S}$}%
\nomenclature[1mb]{$\mathcal{M}$}{influence matrix}%
\nomenclature[1mc]{$\mathcal{M}^t$}{transformed influence matrix}%
\nomenclature[1c]{$C_i$}{set of link failures detected by the sensor $i$}
\nomenclature[1c]{$C^t_i$}{set of pair-wise link failures detected by the sensor $i$}%
\nomenclature[1c]{$\mathcal{C}$}{collection of all $C_i$'s}%
\nomenclature[1c]{$\mathcal{C}^t$}{collection of all $C^t_i$'s}%
\nomenclature[1k]{$k$}{maximum number of link failures detected by any sensor}%
\nomenclature[1n]{$n$}{total number of events}%
\nomenclature[1ma]{$m$}{total number of sensors}%
\nomenclature[1la]{$L$}{localization set}%
\nomenclature[1il]{$I_L$}{normalized localization score}%
\nomenclature[1iw]{$I_W$}{number of elements in the largest localization set}%
\nomenclature[1ii]{$I_I$}{normalized identification score}%
\nomenclature[1id]{$I_D$}{normalized detection score}%

\begin{ack}         \vspace{-0.35cm}                    
This work was supported in part by FORCES (Foundations Of Resilient CybEr-Physical Systems), which receives support from the National Science Foundation (NSF award numbers CNS-1238959, CNS-1238962, CNS-1239054, CNS-1239166), the AFRL LABLET - Science of Secure and Resilient Cyber-Physical Systems (Contract ID: FA8750-14-2-0180, SUB 2784-018400), and the NSF CAREER award \#1453126.
\end{ack} \vspace{-0.25cm}

\bibliographystyle{plain}        
\bibliography{fdi}           

\begin{frontmatter}

\title{Sensor placement for fault location identification in water networks: a minimum test cover approach} 
\author[MIT]{Lina Sela Perelman}
\author[Vanderbilt]{Waseem Abbas}
\author[Vanderbilt]{Xenofon Koutsoukos}
\author[MIT]{Saurabh Amin}

\address[MIT]{Massachusetts Institute of Technology}  
\address[Vanderbilt]{Vanderbilt University}             

\textbf{Supporting Information}

\end{frontmatter}

\section{Transient modeling} \label{app:T}

Unsteady state flow in a closed conduit can be described by mass and momentum equations formulated as \cite{wylie1993}:
\begin{equation}
\frac{\partial h}{\partial t} +  \frac{a^2}{gA}\frac{\partial q}{\partial x} = 0
\end{equation}
\begin{equation}
\frac{1}{gA}\frac{\partial q}{\partial t} + \frac{\partial h}{\partial x} + \frac{cq|q|}{2gDA^2} = 0
\end{equation}
where $h$ is the hydraulic head $[m]$, $q$ is the volumetric flow rate $[\frac{m^3}{sec}]$, $g$ is the gravitational acceleration $[\frac{m}{sec^2}]$, $x$ is distance along the pipe $[m]$, $t$ is the time $[sec]$, $a$ is the wave speed in the conduit $[\frac{m}{sec}]$, $c$ is a friction factor, $D$ is the pipe diameter $[m]$, and $A$ is the pipe cross sectional area $[m^2]$. 

The method of characteristics (MOC) is one of the most common numerical techniques used to approximate the solution of the hydraulic transient. Additional techniques used are finite differences and node characteristic method. A detailed derivation of the governing equations and the solution scheme can be found in \cite{Dal,wylie1993}. The MOC transforms partial differential equations into ordinary differential equations that apply along specific lines (characteristics), $C^+$ and $C^-$, in the \textit{space-time}, $x\text{-}t$, plane. Two characteristic equations are solved explicitly to compute the head and flow, $h_*, q_*$, at new point in time and space, $(\cdot)_*$, given that the conditions at a previous time step along the characteristic grid are known, i.e., $h_+, q_+$ and $h_-, q_-$. For a given pipe, the two comparability equations are formulated as:
\begin{subequations} \label{eq:a3}
\begin{align} 
		C^+: \frac{a}{gA}\left ( q_* - q_+\right) + \left ( h_* - h_+ \right ) + \frac{c\Delta x}								{2gDA^2}q_+|q_+| = 0 \label{eq:a3a} \\ 
		C^-: \frac{a}{gA}\left ( q_* - q_-\right) - \left ( h_* - h_- \right ) + \frac{c\Delta x}							{2gDA^2}q_-|q_-| = 0 \label{eq:a3b}
\end{align} 
\end{subequations}
Rearranging equations \eqref{eq:a3a} and \eqref{eq:a3b} we get:
\begin{subequations} \label{eq:a4}
\begin{align} 
		C^+: h_* = C_P - bq_* \label{eq:a4a} \\ 
		C^-: h_* = C_M + bq_*  \label{eq:a4b}
\end{align} 
\end{subequations}
where 
\begin{subequations}
\begin{align} 
		C^+: & \hspace{0.5em} C_P = h_+ + q_+ \left ( b - r |q_+| \right ) \\
		C^-:  &\hspace{0.5em} C_M = h_- - q_- \left ( b - r |q_-| \right ) 
\end{align} 
\end{subequations}
and 
\begin{equation}
b = \frac{a}{gA}
\end{equation}
\begin{equation}
r = \frac{c \Delta x}{2gDA^2}
\end{equation}
$b$ is a function of the physical characteristics of the pipe and the wave speed of the fluid in the conduit. The parameter $b$ can be viewed as the characteristic impedance, which is associated with the transient state. $r$ is a function of the physical characteristics of the pipe, that can be viewed as pipe's resistance coefficient, and is associated with the steady state. If $b=0$ the set of equations \eqref{eq:a4} is reduced to the steady state equations, where the head losses along the pipe occur only due to friction.

We designate the points $(\cdot)_+,(\cdot)_-,(\cdot)_*$ over a \textit{space-time} grid of characteristics. If $i$ and $t$ are indices for space and time, respectively, then: $(\cdot)_* \rightarrow (h_{i,t+1},q_{i,t+1}),(\cdot)_+ \rightarrow (h_{i-1,t},q_{i-1,t}),(\cdot)_- \rightarrow (h_{i+1,t},q_{i+1,t})$. Then solving first for $h_{i,t+1}$, by eliminating $q_*$ in \eqref{eq:a4}, for a single node in the numerical grid, we get:
\begin{align} 
 h_{i,t+1} &= \frac{1}{2} \big[ h_{i-1,t} + h_{i+1,t} + b \left( q_{i-1,t} - q_{i+1,t} \right) \nonumber \\
 &\qquad {} + r \left( q_{i+1,t} |q_{i+1,t}| - q_{i-1,t} |q_{i-1,t}| \right) \big] \label{eq:3a}\\
q_{i,t+1} &= \frac{1}{b} \big[ h_{i,t+1} -h_{i+1,t} + q_{i+1,t} - r |q_{i+1,t}| \big]  \label{eq:3b}
\end{align} 
where $r$ is the resistance coefficient, which is associated with the steady state, and $b$ is the impedance coefficient, which is associated with the transient state. If $b = 0$ the set of equations \eqref{eq:3a},\eqref{eq:3b} is reduced to the steady state, where the head loss along a pipe occurs only due to friction \cite{todi}.

At the boundaries specific conditions need to be defined describing the head-flow relation. Common boundary conditions, such as cross-connections and control valves, can be found in \cite{wylie1993}. We give an example for boundary condition for pipe burst at location $i$ using the orifice head-flow equation:
\begin{equation} \label{eq:5a}
h_{i,t+1}  + \frac{b}{2}C_dA_{d,t+1} \sqrt{2gh_{i,t+1}}
- \frac{C_M +C_P }{2} = 0
\end{equation}
where $C_d$ is the orifice discharge coefficient, $A_{d}$ is the cross-section area of the orifice, $C_P = h_{i-1,t} + q_{i-1,t} \left ( b - r |q_{i-1,t}| \right ) $, $ C_M = h_{i+1,t} - q_{i+1,t} \left ( b - r |q_{i+1,t}| \right )$. Before the burst occurs the coefficient $A_d$ is equal to zero and Equation\eqref{eq:5a} reduces to Equation \eqref{eq:3a}. During a burst $A_d$ is positive, hence we can expect a change in the hydraulic head. The relationship between the head and the pressure measured by sensors at location $i$ is relative to the elevation of location $i$, denoted by $z_i$, i.e., at any given time, $p_{i,t} = \left( h_{i,t} - z_i \right)\rho g$.  
Hence, we can expect to detect the pipe burst by observing the differences between the expected and the measured pressures at a given time and location in the network. Similar approaches have been previously suggested in \cite{69012}.

\begin{figure}
        \centering
                \includegraphics[scale=0.3]{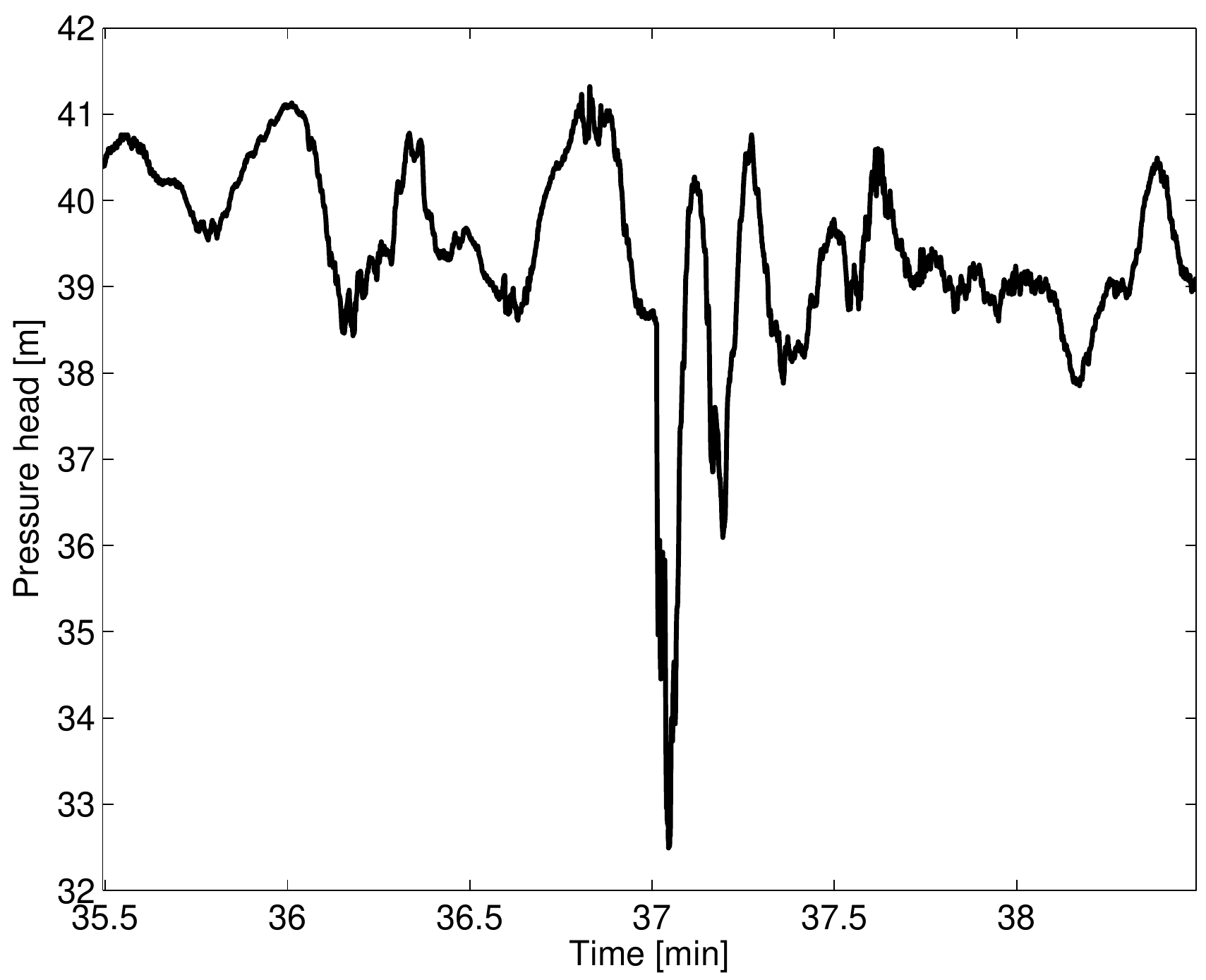}
       \caption{Pressure signal during a burst event recorded from online sensor installed in a water system}  \label{fig:2}
\end{figure}

Figure \ref{fig:2} shows a raw pressure signal recorded by Visenti \cite{visent} online sensor during a pipe burst event with $250 [Hz]$ sampling frequency. Figure \ref{fig:2} shows the dynamic nature of pressure, a sharp drop in the pressure during a pipe burst event, and a rapid return to normal operating range. The duration of drop in pressures is just under a few seconds, hence cannot be detected using a more traditional methods such as supervisory control and data acquisition (SCADA) systems, which typically operate on minutes scales.


\section{Submodularity} \label{app:A}
\begin{lemma}
\label{lem:submodular}
The detection function $f_D$ (as defined in the Equation (6) of the main paper) is submodular.
\end{lemma}
\textit{Proof -- }
Let $\mathcal{C}_s\subseteq \mathcal{C}_r\subseteq \mathcal{C}$, and $C_i\in\mathcal{C}\setminus \mathcal{C}_r$, then we need to show 
$$f_D\left(\mathcal{C}_s\cup\{C_i\}\right) - f_D(\mathcal{C}_s) \ge f_D\left(\mathcal{C}_r\cup\{C_i\}\right) - f_D(\mathcal{C}_r)$$
Assume that $C_i'=C_i\setminus\bigcup\limits_{C_j\in \mathcal{C}_s}C_j$, then
\begin{equation}
\label{eq:p1}
f_D(\mathcal{C}_s\cup\{C_i\}) = f_D(\mathcal{C}_s\cup\{C_i'\}) = f_D(\mathcal{C}_s) + f_D(\{C_i'\})
\end{equation}
Moreover, let $\lambda = \left(\bigcup\limits_{C_k\in \mathcal{C}_r}C_k\right)\setminus \left(\bigcup\limits_{C_j\in \mathcal{C}_s}C_j \cup C_i'\right)$, and $\mu = \bigcup\limits_{C_k\in \mathcal{C}_r}C_k\cap C_i'$, then
\begin{equation}
\label{eq:p2}
f_D(\mathcal{C}_r \cup\{C_i\}) = f_D(\mathcal{C}_r\cup\{C_i'\}) = f_D(\mathcal{C}_s\cup\{C_i'\}) + f_D(\{\lambda\}),
\end{equation}
and
\begin{equation}
\label{eq:p3}
f_D(\mathcal{C}_r) = f_D(\mathcal{C}_s) + f_D(\{\lambda\}) +f_D(\{\mu\}).
\end{equation}

Substituting \eqref{eq:p3} into \eqref{eq:p2} gives,
\begin{equation*}
f_D(\mathcal{C}_s\cup\{C_i\}) -  f_D(\mathcal{C}_s) -f_D(\{\mu\}) = f_D(\mathcal{C}_r \cup\{C_i\}) - f_D(\mathcal{C}_r) 
\end{equation*}
The required result follows directly. \qed

\section{Augmented greedy -- Example 3 (cont.)}
In each iteration, for every sensor $i$ not in the test cover, $C_i$ is decomposed into two sets namely, $X_i=C_i\setminus\mathcal{C}_{cov}$ and $Y_i=C_i\cap\mathcal{C}_{cov}$. The utility of including a sensor in the test cover is calculated in terms of $x_i$ and $y_i$. $x_i$ computes the number of pair-wise link failures detected by $C_i$ corresponding to the links not in $\mathcal{C}_{cov}$, whereas $y_i$ computes the undetected pair-wise link failures corresponding to the links in $\mathcal{C}_{cov}$ that can be detected by $C_i$. Then, a sensor that maximizes the utility is selected and $\mathcal{C}_{cov}$, which is the set of covered (detected) events, and $G_u$, which is the set of undetected pair-wise events corresponding to the events detected by the sensor $u$ already included in the test cover, are updated. We give detailed steps of the algorithm using the illustrative example in the paper (Figure \ref{fig:3}). 

\begin{figure}[ht]
\centering
\includegraphics[scale = 0.5]{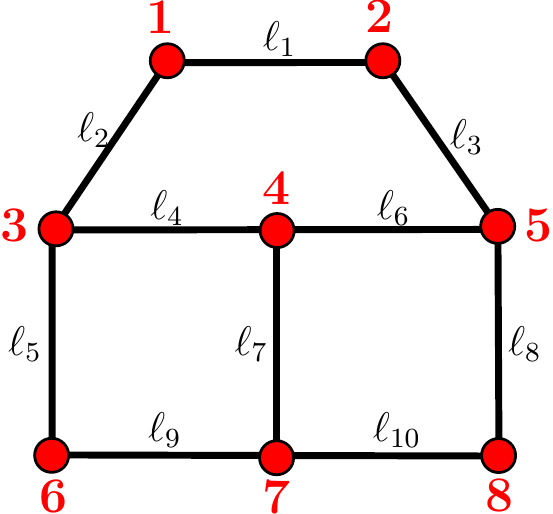}
\caption{Illustrative example layout}
\label{fig:3}
\end{figure}
Recall the influence matrix: \vspace{-0.5cm}
\footnotesize
$$\mathcal{M}( \mathcal{L},\mathcal{S}) = 
\bordermatrix{
& S_1& S_2& S_3& S_4& S_5& S_6& S_7& S_8\cr
\ell_1&1 & 1 & 1 & 0 & 1 & 0 & 0 & 0 \cr 
\ell_2&1 & 1 & 1 & 1 & 0 & 1 & 0 & 0 \cr
\ell_3&1 & 1 & 0 & 1 & 1 & 0 & 0 & 1 \cr 
\ell_4&1 & 0 & 1 & 1 & 1 & 1 & 1 & 0 \cr
\ell_5&1 & 0 & 1 & 1 & 0 & 1 & 1 & 0 \cr
\ell_6&0 & 1 & 1 & 1 & 1 & 0 & 1 & 1 \cr 
\ell_7&0 & 0 & 1 & 1 & 1 & 1 & 1 & 1 \cr
\ell_8&0 & 1 & 0 & 1 & 1 & 0 & 1 & 1 \cr
\ell_9&0 & 0 & 1 & 1 & 0 & 1 & 1 & 1 \cr
\ell_{10}& 0 & 0 & 0 & 1 & 1 & 1 & 1 & 1}
$$

\normalsize
\begin{table*}[t] \scriptsize
\tabcolsep=0.11cm
	\renewcommand{\arraystretch}{1.1} 
\centering
  \caption{Illustrative example demonstrating the steps in the augmented greedy solution of the MTC problem} \label{app:tab}
    \begin{tabular}{|l|l|l|l|l|l|}
    \hline
     ~   & $j=1$ & $j=2$ & $j=3$ & $j=4$ & $j=5$\\ \hline   
   $\mathcal{C}_{cov}$ & $\emptyset$ & $\{1,2,3,4,5\}$ & $\{1,2,3,4,5,6,8\}$ & $\{1,2,\cdots,9\}$ & $\{1,2,\cdots,10\}$\\ \hline 
   $n_j$ & $10$ & $5$ & $3$ & $1$ & $0$\\ \specialrule{.1em}{.1em}{.1em}
   $X_1,Y_1$ & $\mathbf{\{1,2,3,4,5\}},\;\emptyset$ & -- & -- & -- & -- \\ \hline
   $X_2,Y_2$ & $\{1,2,3,6,8\},\;\emptyset$ & $\mathbf{\{6,8\},\{1,2,3\}}$ & -- & --& --\\ \hline
   $X_3,Y_3$ & $\{1,2,4,5,6,7,9\},\;\emptyset$ & $\{6,7,9\},\{1,2,4,5\}$ & $\mathbf{\{7,9\},\{1,2,4,5,6\}}$ & -- & --\\ \hline
   $X_4,Y4$ & $\{2,3,\cdots,10\},\;\emptyset$ & $\{6,\cdots,10\},\{2,3,4,5\}$ & $\{7,9,10\},\{2,\cdots,6,8\}$ & $\{10\},\{2,3,\cdots,9\}$ & $\emptyset,\;\{2,3,\cdots,10\}$\\ \hline
   $X_5,Y_5$ & $\{1,3,4,6,7,8,10\},\;\emptyset$ & $\{6,7,8,10\},\{1,3,4\}$ & $\{7,10\},\{1,3,4,6,8\}$ & $\mathbf{\{10\},\{1,3,4,6,7,8\}}$ & --\\ \hline
   $X_6,Y_6$ & $\{2,4,5,7,9,10\},\;\emptyset$ & $\{7,9,10\},\{2,4,5\}$ & $\{7,9,10\},\{2,4,5\}$ &$\{10\},\{2,4,5,7,9\}$ & $\emptyset,\;\{2,4,5,7,9,10\}$\\ \hline
   $X_7,Y_7$ & $\{4,5,\cdots,10\},\;\emptyset$ & $\{6,\cdots,10\},\{4,5\}$ & $\{7,9,10\},\{4,5,6,8\}$ & $\{10\},\{4,5,\cdots,9\}$ & $\emptyset,\;\{4,5,\cdots,10\}$\\ \hline
   $X_8,Y_8$ & $\{3,6,\cdots,10\},\;\emptyset$ & $\{6,\cdots,10\},\{3\}$ & $\{7,9,10\},\{3,6,8\}$ & $\{10\},\{3,6,7,8,9\}$ & $\emptyset,\;\{3,6,\cdots,10\}$\\ \specialrule{.1em}{.1em}{.1em} 
   $x_1,\;y_1$ & $\mathbf{25,0}^\ast$ & -- & -- & -- & --\\ \hline
   $x_2,\;y_2$ & $25,0$ & $\mathbf{6,6}^\ast$ & --& -- & --\\ \hline
   $x_3,\;y_3$ & $21,0$ & $6,4$ & $\mathbf{2,3}^\ast$& -- & --\\ \hline
   $x_4,\;y_4$ & $9,0$ & $0,4$ & $0,2$ & $0,1$ & $0,0$\\ \hline
   $x_5,\;y_5$ & $21,0$ & $4,6$ & $2,3$ & $\mathbf{0,3}^\ast$ & -- \\ \hline
   $x_6,\;y_6$ & $24,0$ & $6,6$ & $0,2$ & $0,1$ & $0,0$\\ \hline
   $x_7,\;y_7$ & $21,0$ & $0,6$ & $0,0$ & $0,0$ & $0,0$\\ \hline
   $x_8,\;y_8$ & $24,0$ & $0,4$ & $0,2$ & $0,0$ & $0,0$\\ \specialrule{.1em}{.1em}{.1em} 
   $G_0$ & $\emptyset$ & $\emptyset$ & $\emptyset$ & $\emptyset$ & $\emptyset$ \\ \hline
   $G_1$ & $\left\{
\begin{array}{lll}
\{1,2\},\{1,3\},\{1,4\},\\
\{1,5\},\{2,3\},\{2,4\},\\
\{2,5\},\{3,4\},\{3,5\},\\
\{4,5\}\\
\end{array}  
\right\}$ & $\left\{
\begin{array}{lll}\{1,2\},\{1,3\},\{2,3\},\\
\{4,5\}\\
\end{array}\right\}$ & $\{\{1,2\},\{4,5\}\}$ & $\emptyset$ & $\emptyset$ \\ \hline
   $G_2$ & --  & $\{\{6,8\}\}$ & $\emptyset$ & $\emptyset$ & $\emptyset$\\ \hline
   $G_3$ & -- & -- & $\{\{7,9\}\}$ & $\emptyset$& $\emptyset$ \\ \hline
   $G_4$ & -- & -- & --& $\{10\}\rightarrow\emptyset$ & $\emptyset$\\ \hline
   
    \end{tabular}\\
$^\ast$ is the selected sensor with the maximum utility, i.e. $w_{i^{\ast}}\gets \max w_i$.
\end{table*}
\begin{description}
\item[Initialization] $\mathcal{C}_{cov} = \emptyset$; $\mathcal{C}^{{\ast}}=\emptyset $; $G_0=\emptyset$; $n = 10$;
\item[First iteration of the while loop, $j=1$.] We denote the total number of events detected by sensor $i$ as $k_i$, i,.e., $k_i = |C_i|$. Similarly, $k_{i,j}$ denotes the number of undetected events that are detected by the sensor $i$ in the $j^{th}$ iteration, i.e., $k_{i,j} = |C_i\setminus \mathcal{C}_{cov}|$. In the first iteration $k_{i,j} = k_i$, $\forall i$. In this example, the set of all $k_{i,1}$'s is $\{5,5,7,9,7,6,7,6\}$.
 Then, for each sensor $i$, we compute the number of new pair-wise events detected, $x_i = k_{i,1}(n - k_{i,1})$. For instance, for sensor $1$, $x_1=5(10-5)=25$. Next, we need to compute $y_i$ for all $i$. Since there is no sensor in the test cover in the first iteration, $y_i=0$ for all $i$. The total utility of selecting a sensor is equal to $w_i = x_i + y_i$. The maximum $w_{i^\ast}$ is attained for sensors $1$ and $2$. We select sensor $1$ to be included in the test cover, and update $\mathcal{C}^{{\ast}}\gets \{C_{1^{\ast}}\}$, and $G_1$, which is the set of all undetected pair-wise events corresponding to the events in $X_1 = C_1\setminus \mathcal{C}_{cov}$. Here, $G_1 =  \left\{\{1,2\},\{1,3\},\cdots,\{4,5\} \right\}$. Finally, we update the set of covered (detected) events $\mathcal{C}_{cov}~\gets~\mathcal{C}_{cov}\cup C_1 = C_1=\{1,2,3,4,5\}$. 

\item[Second iteration of the while loop, $j=2$.] At the beginning of second iteration, the event space has been reduced from 10 to 5, i.e., $n_2 = 5$. For each sensor $i$, we first compute $X_i$, which is the set of undetected events (events that are not in $\mathcal{C}_{cov}$) that are detected by the sensor $i$, i.e., $X_i \gets \left(C_i \setminus \mathcal{C}_{cov}\right)$. Then, we compute $x_i = k_{i,2}(n_2 - k_{i,2})$, where $k_{i,2}$ is $| X_i|$. For instance, for sensor $2$, $C_2 = \{1,2,3,6,8 \}$, then $X_2 \gets \left(C_2 \setminus \mathcal{C}_{cov}\right) = \{6,8\}$ and $k_{2,2} = 2$. Then $x_2 = 2(5-2) = 6$. Next, for each sensor $i$, we compute $y_i$, which is the number of pair-wise events in $G_1$ that are detected by the sensor $i$. For instance, in the case of sensor $2$, six of the pair-wise events in $G_1$, given by $\{\{1,4\},\{1,5\},\{2,4\},\{2,5\},\{3,4\},\{3,5\}\}$, are detected by the sensor $2$. Thus, we get $y_2 = 6$. The values of $y_i$ for all $i$ are given in Table 1. After this, the utility of each sensor is computed as $w_i = x_i + y_i$. For sensor $2$, the value of $w_2$ is $12$, which turns out to be the maximum among all the sensors in the second iteration. Thus, sensor $2$ is included in the test cover. 
We update $\mathcal{C}^{{\ast}}\gets \mathcal{C}^{{\ast}} \cup \{C_{2^{\ast}}\}$, $\mathcal{C}_{cov}= \{1,2,3,4,5,6,8\}$, and 
\begin{equation*}
\begin{split}
G_1 &\gets G_1 \setminus   \left\{\{1,4\},\{1,5\},\{2,4\},\{2,5\},\{3,4\},\{3,5\} \right\}\\
&= \{\{1,2\},\{1,3\},\{2,3\},\{4,5\}\}.
\end{split}
\end{equation*}
At the same time, a new set $G_2$ is created, which contains the set of pair-wise events in $X_2$. Since $X_2 = \{6,8\}$, we get $G_2 = \{\{6,8\}\}$.

\item[Next iteration.] We continue with the same steps until no improvement can be made, i.e. $w_i = 0$ for each sensor. At the end of the algorithm, sensors in the set $\{1,2,3,5\}$ are included in the test cover.
\end{description}

For this example, a complete account of the values of variables in each iteration of the algorithm is given in Table 1.

\section{Evaluation on real networks (cont.)}
For all networks \cite{datab,doi:10.1061/(ASCE)WR.1943-5452}, the layouts and the simulation plots illustrating the four performance metrics are shown in Table \ref{tab:10}. For the ease of presentation, the worst localization set size, $I_W$, is normalized by dividing it by the number of pipes.

\centering
\begin{table*}[ht]
\caption{Evaluation on real netowrks}
\centering
\begin{tabular}{|c|c|}
\hline
\includegraphics[trim = 10mm 60mm 10mm 60mm, clip, scale = 0.25]{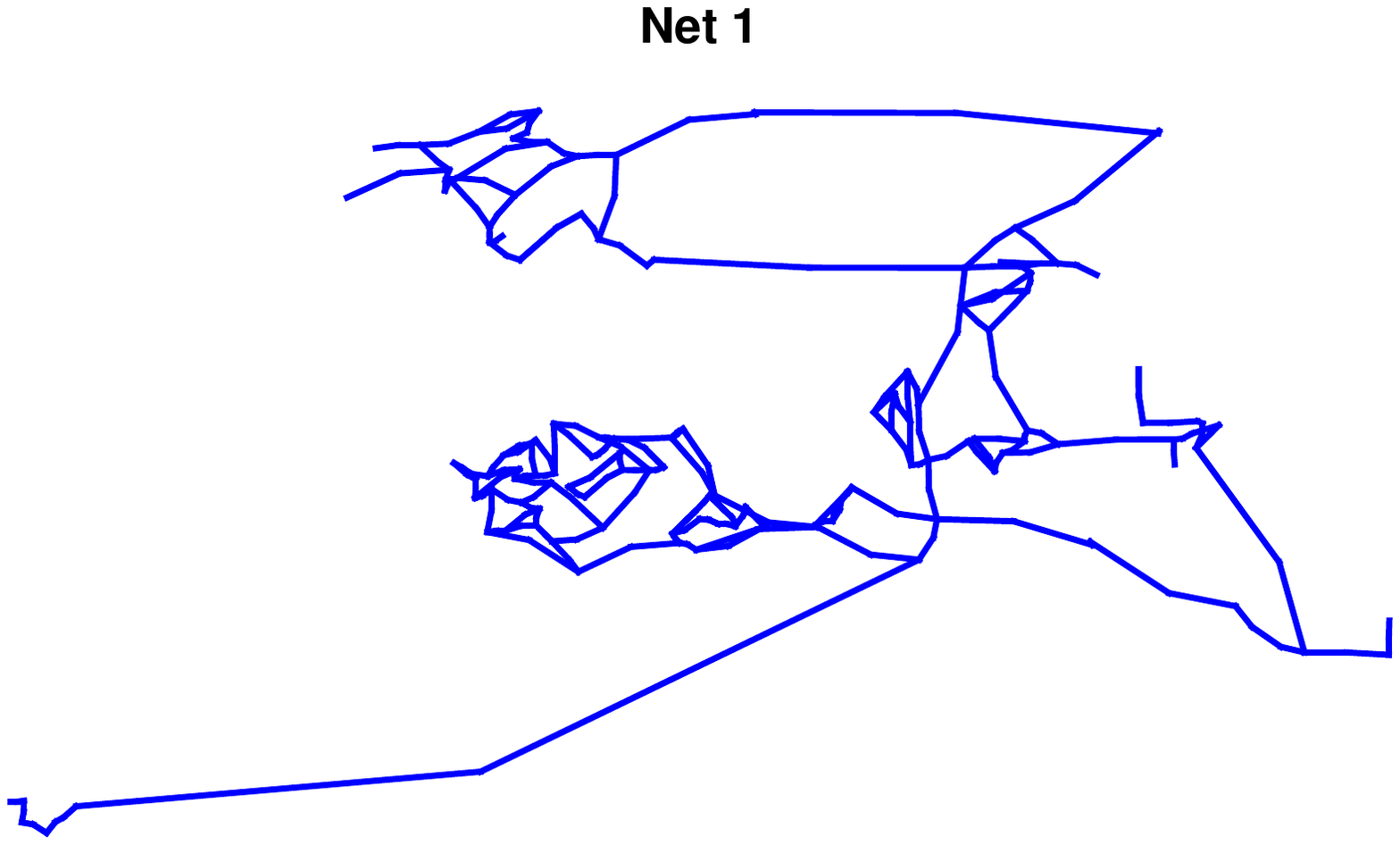}&\includegraphics[trim = 10mm 60mm 10mm 60mm, clip, scale = 0.25]{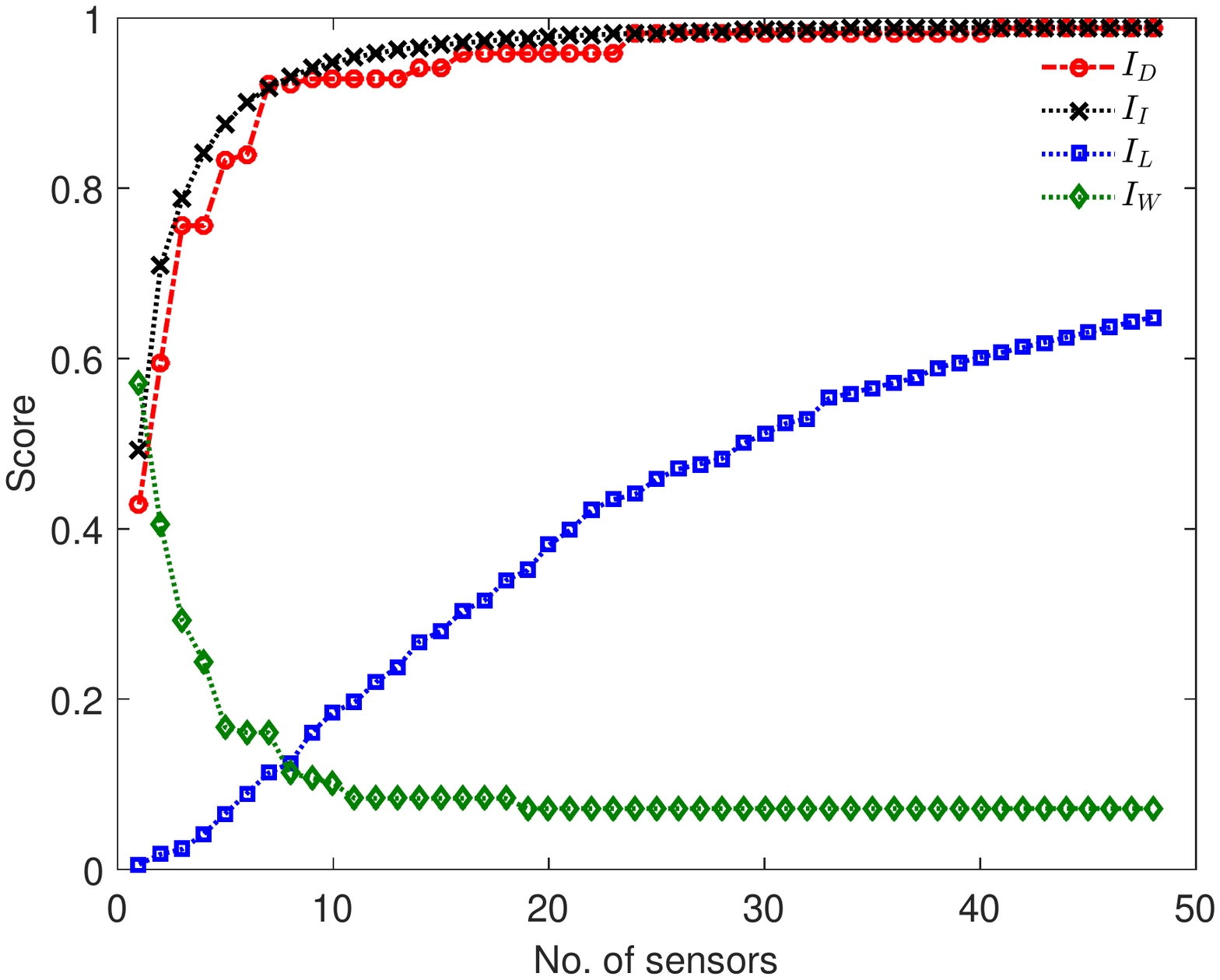}\\
 \hline
\includegraphics[trim = 10mm 60mm 10mm 60mm, clip, scale = 0.25]{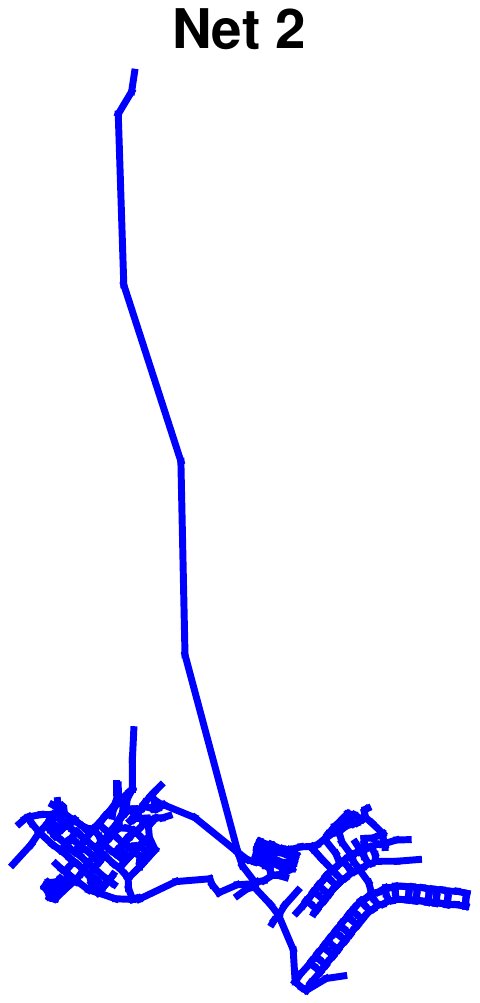}&\includegraphics[trim = 10mm 60mm 10mm 60mm, clip, scale = 0.25]{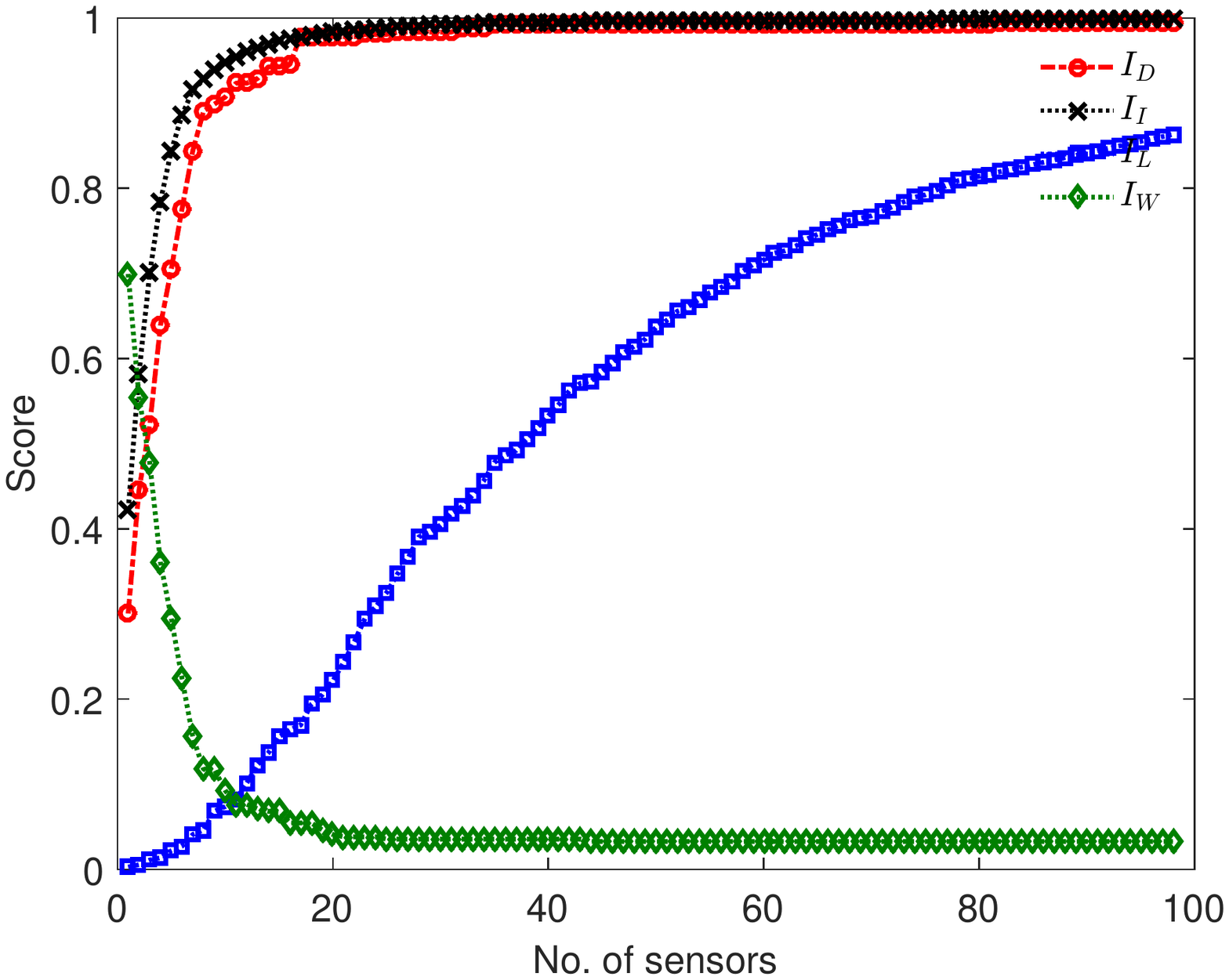}\\
\hline
\includegraphics[trim = 10mm 60mm 10mm 60mm, clip, scale = 0.25]{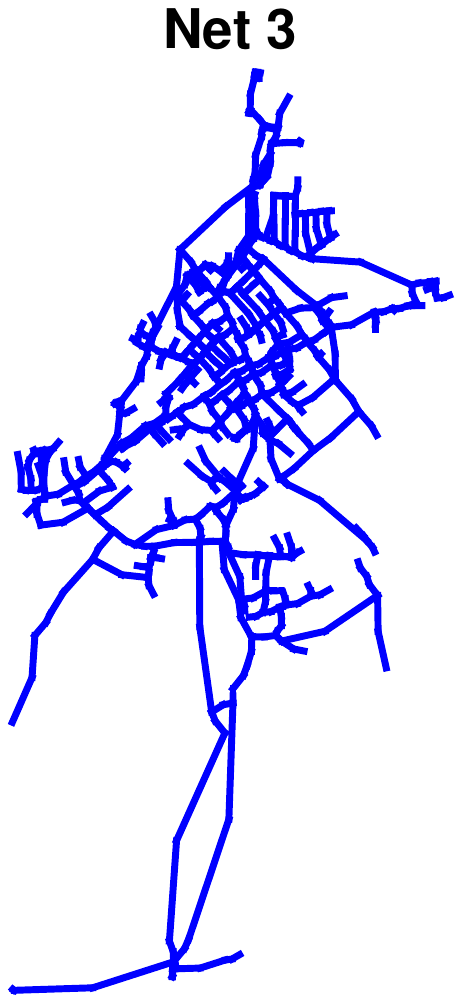}&\includegraphics[trim = 10mm 60mm 10mm 60mm, clip, scale = 0.25]{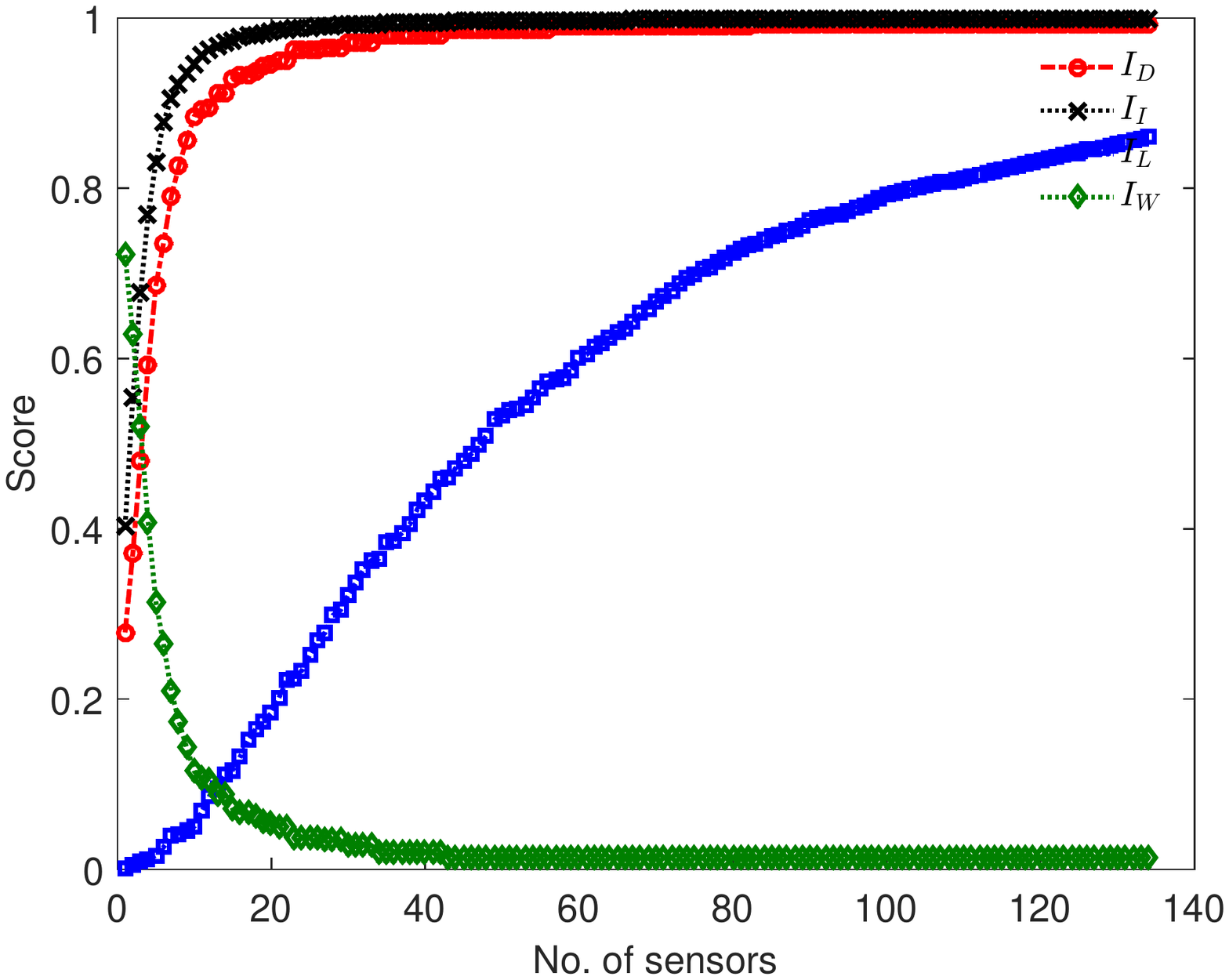}\\
 \hline
\includegraphics[trim = 10mm 60mm 10mm 60mm, clip, scale = 0.25]{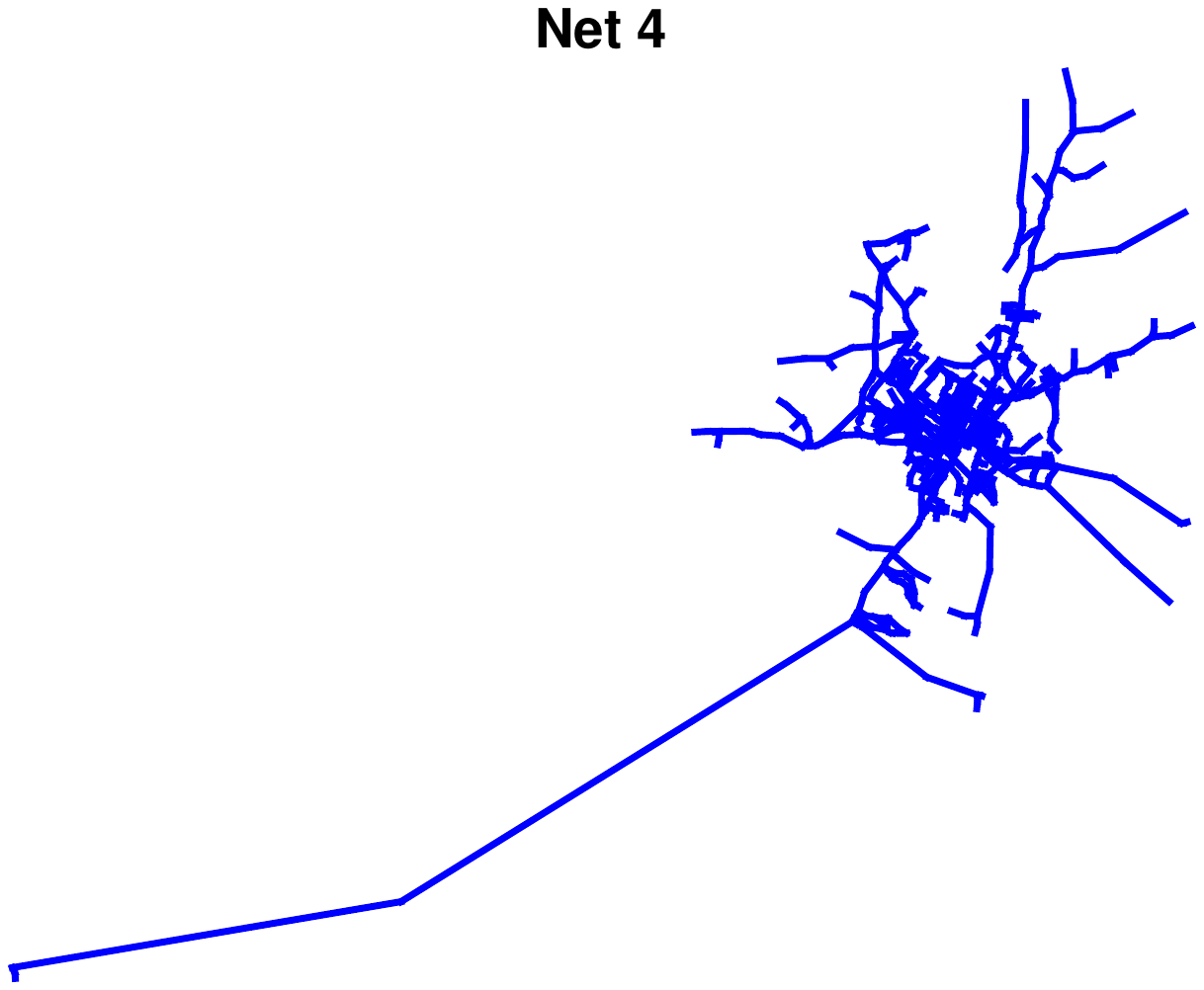}&\includegraphics[trim = 10mm 60mm 10mm 60mm, clip, scale = 0.25]{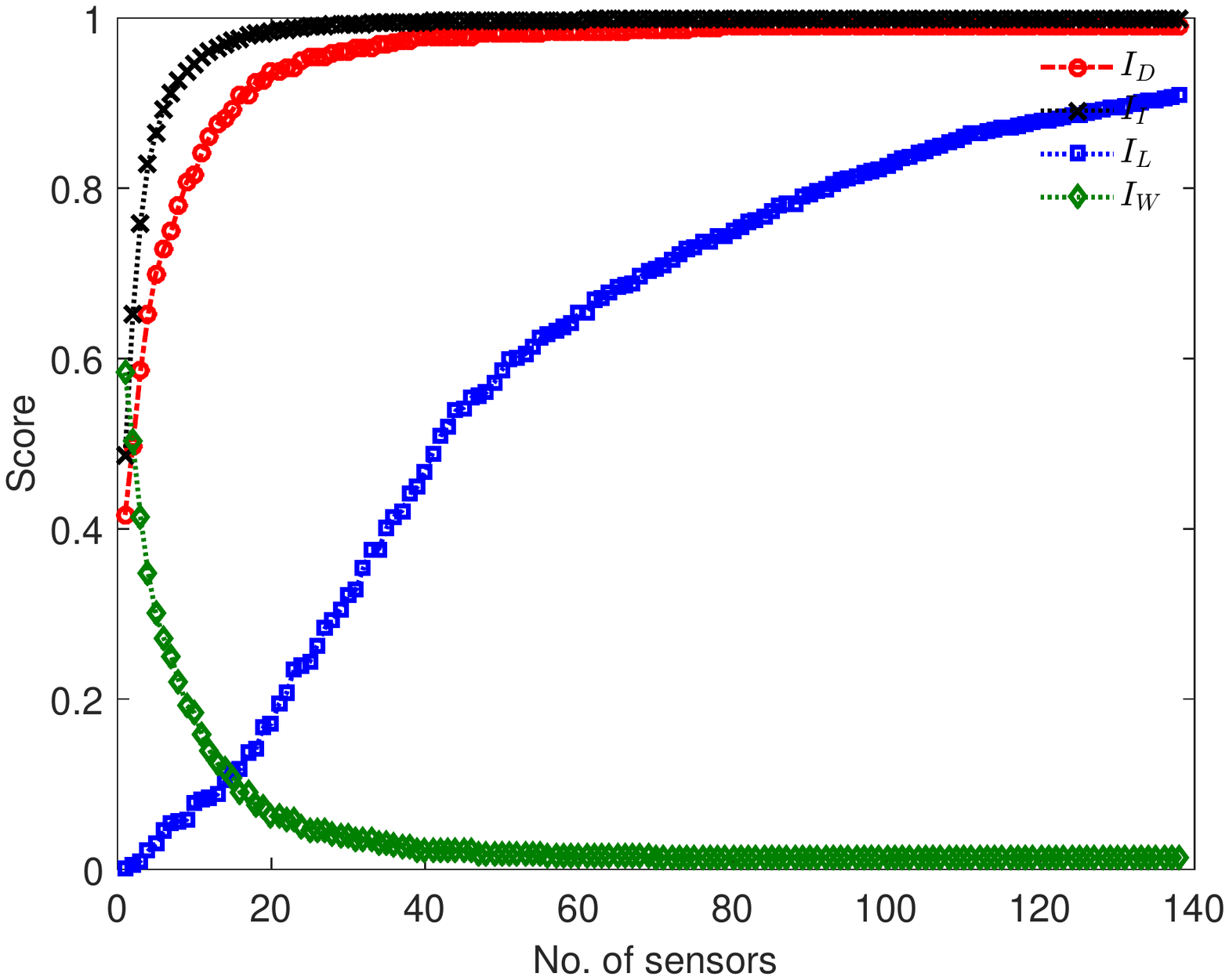}\\
\hline
\includegraphics[trim = 10mm 60mm 10mm 60mm, clip, scale = 0.25]{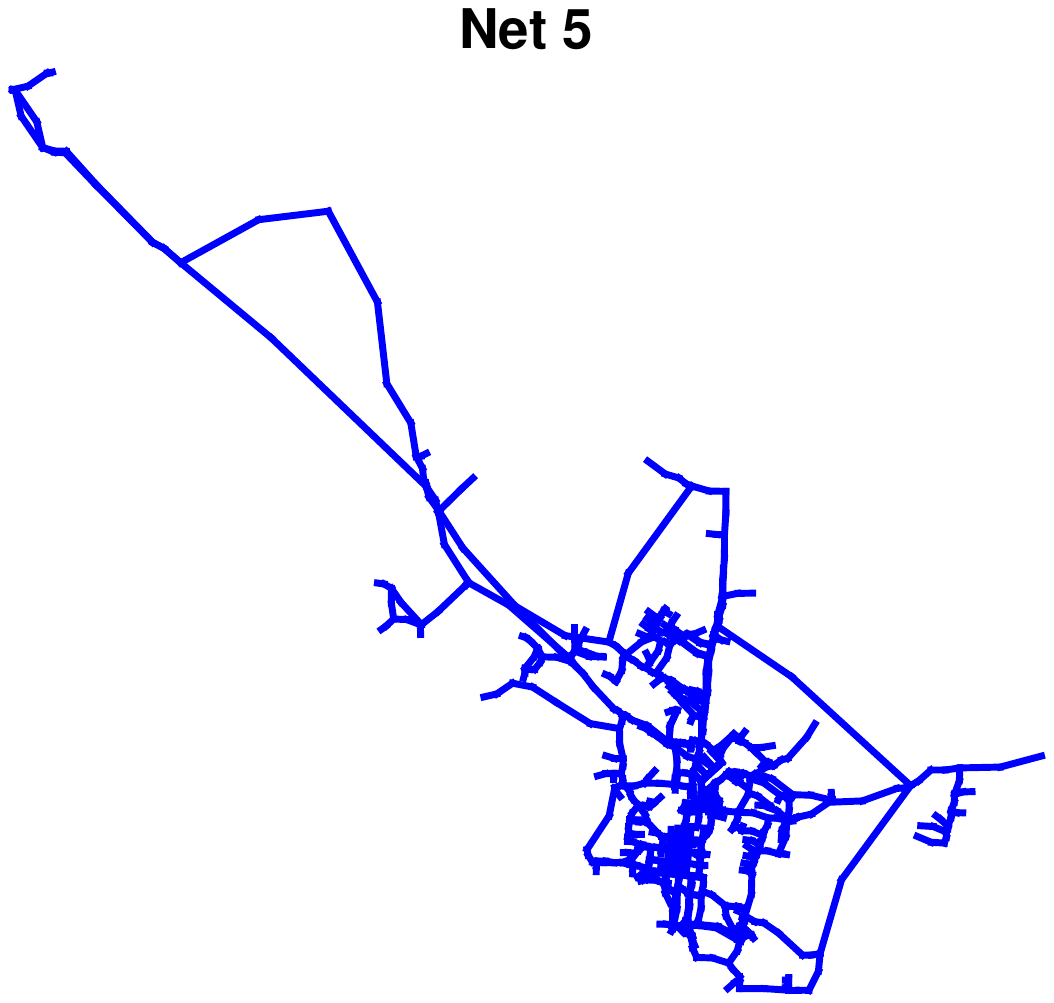}&\includegraphics[trim = 10mm 60mm 10mm 60mm, clip, scale = 0.25]{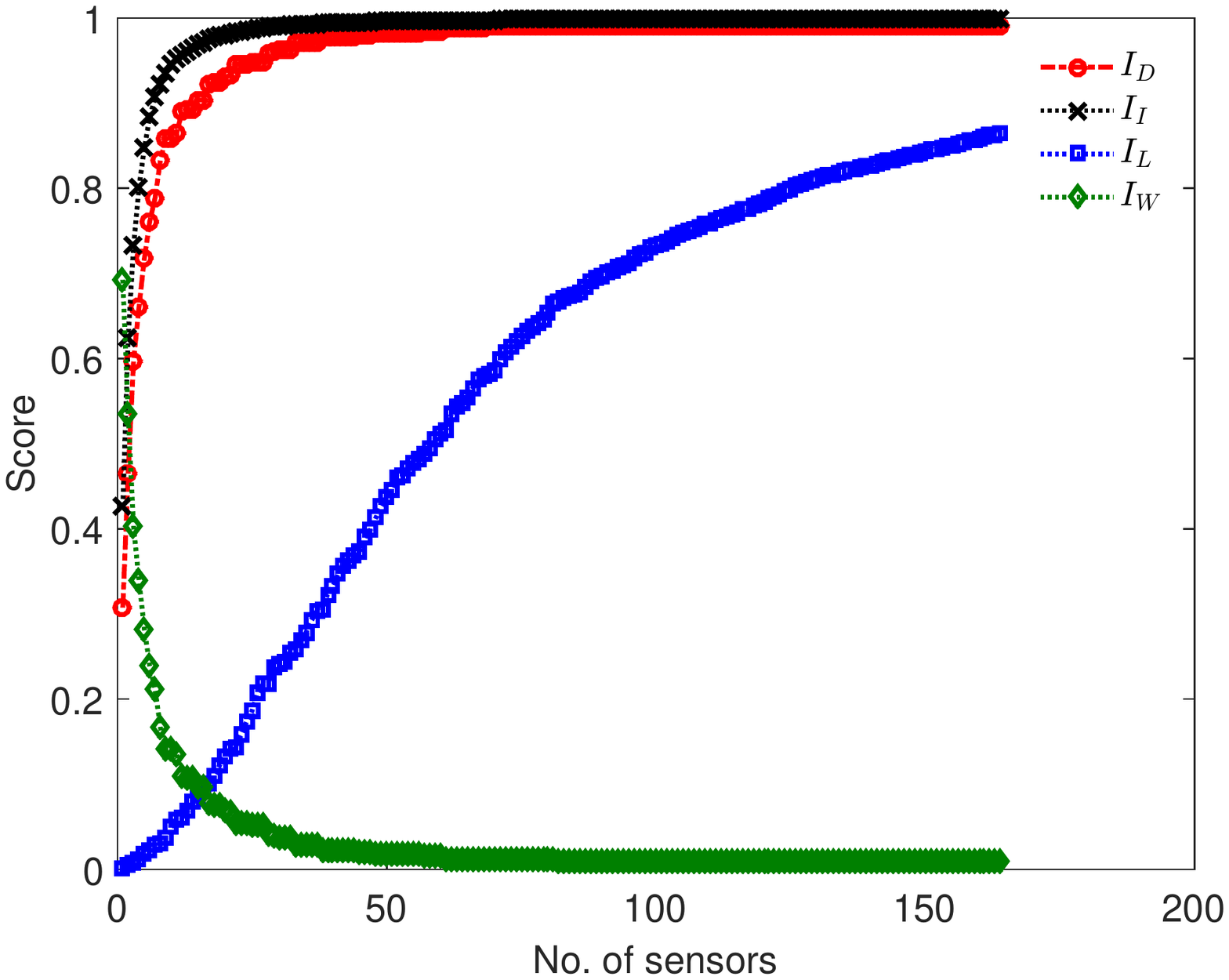}\\
\hline

\end{tabular}
\label{tab:10}
\end{table*}

\begin{table*}[ht]
\caption*{Evaluation on real netowrks}
\centering
\begin{tabular}{|c|c|}
 \hline
\includegraphics[trim = 10mm 60mm 10mm 60mm, clip, scale = 0.25]{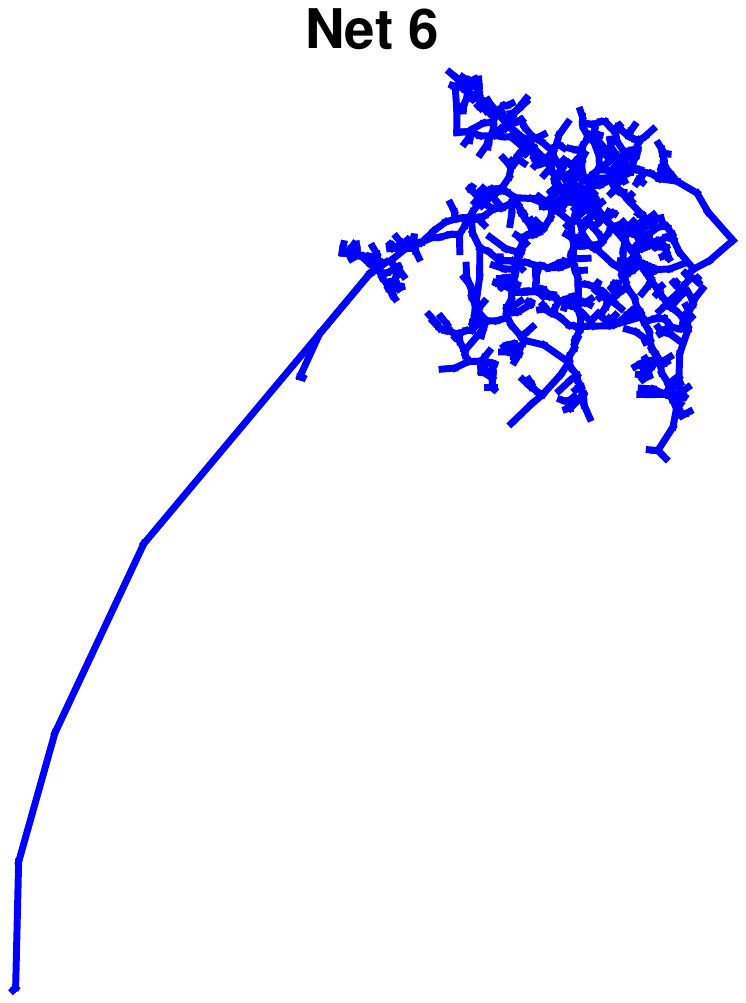}&\includegraphics[trim = 10mm 60mm 10mm 60mm, clip, scale = 0.25]{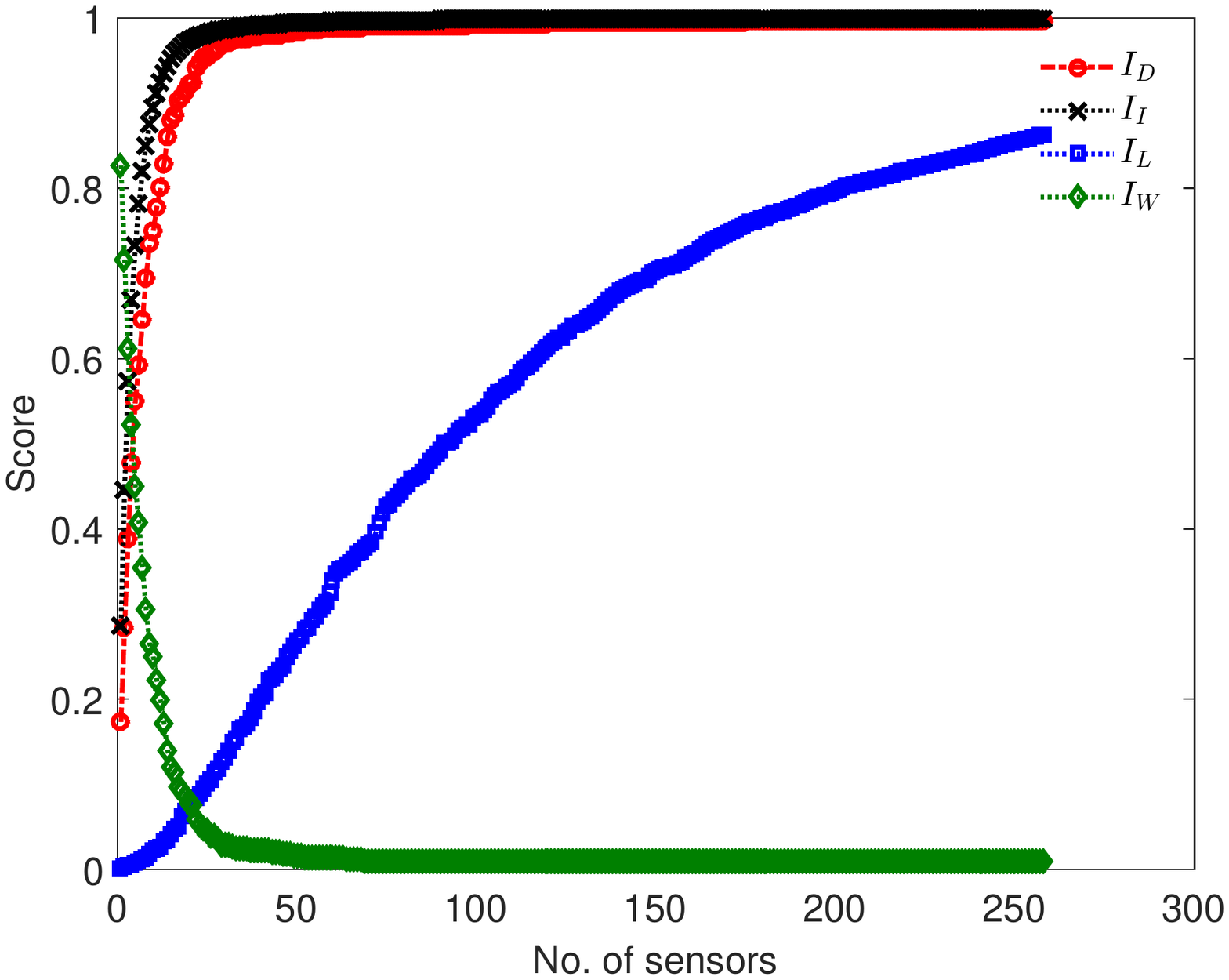}\\
\hline
\includegraphics[trim = 10mm 60mm 10mm 60mm, clip, scale = 0.25]{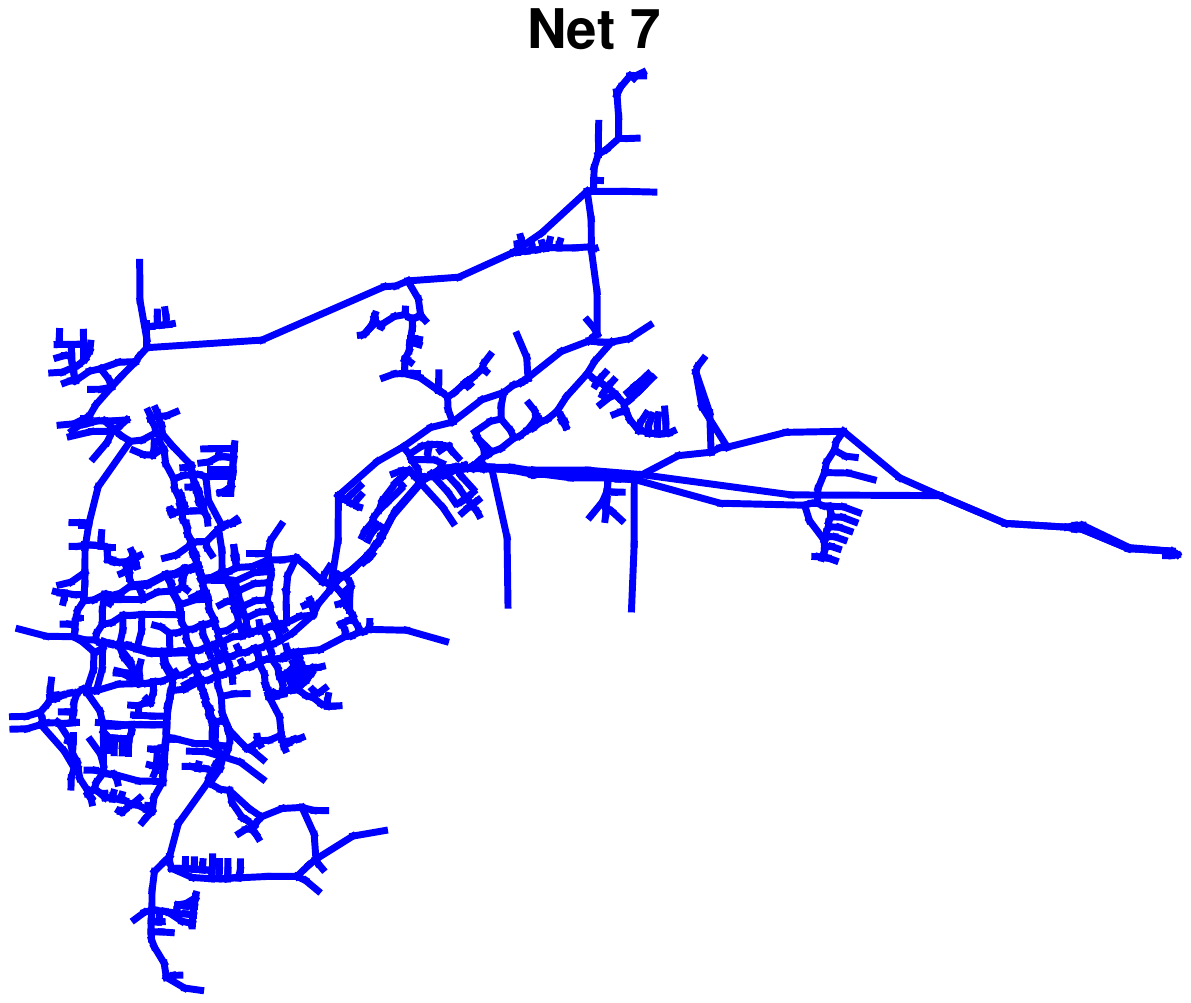}&\includegraphics[trim = 10mm 60mm 10mm 60mm, clip, scale = 0.25]{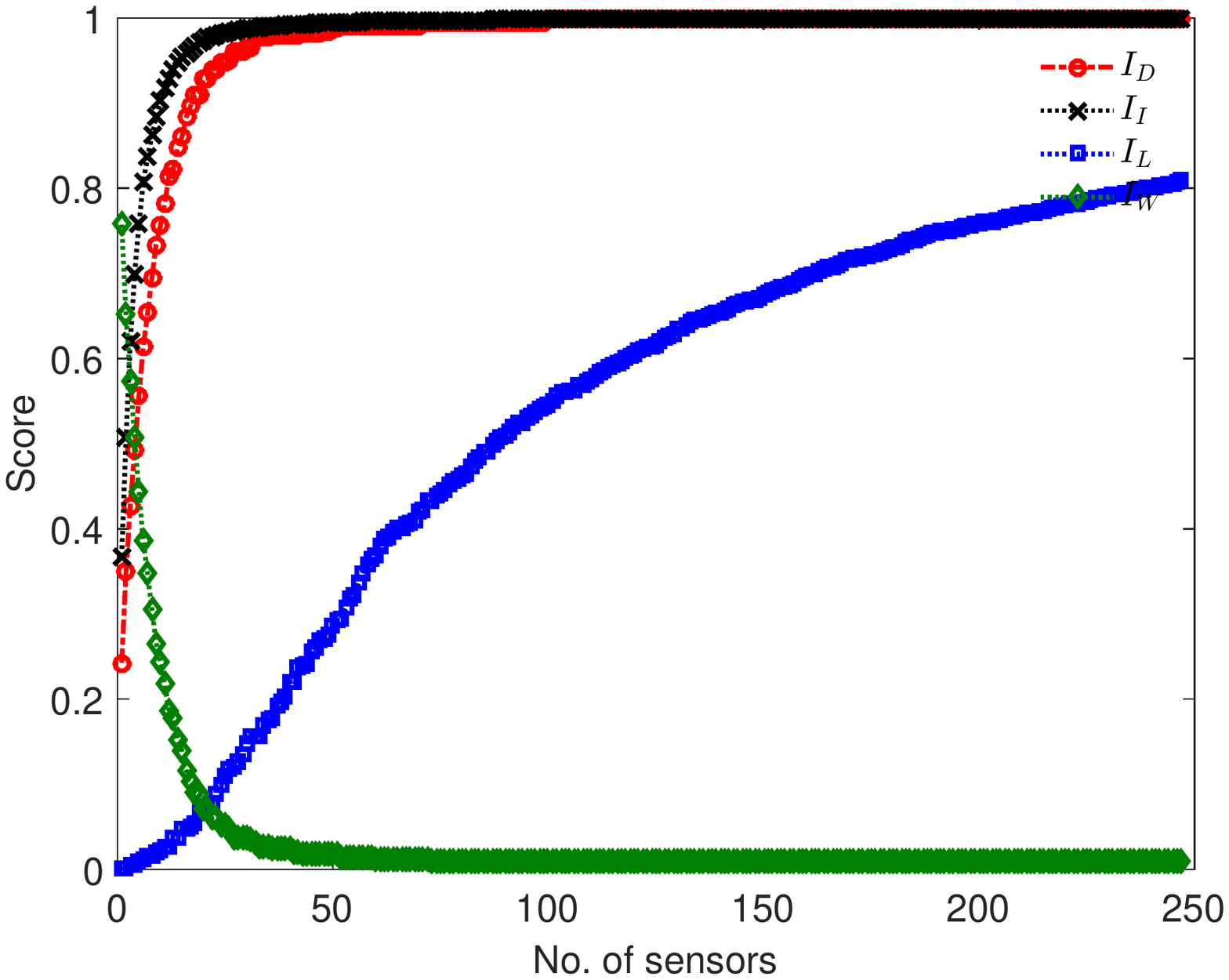}\\
 \hline
\includegraphics[trim = 10mm 60mm 10mm 60mm, clip, scale = 0.25]{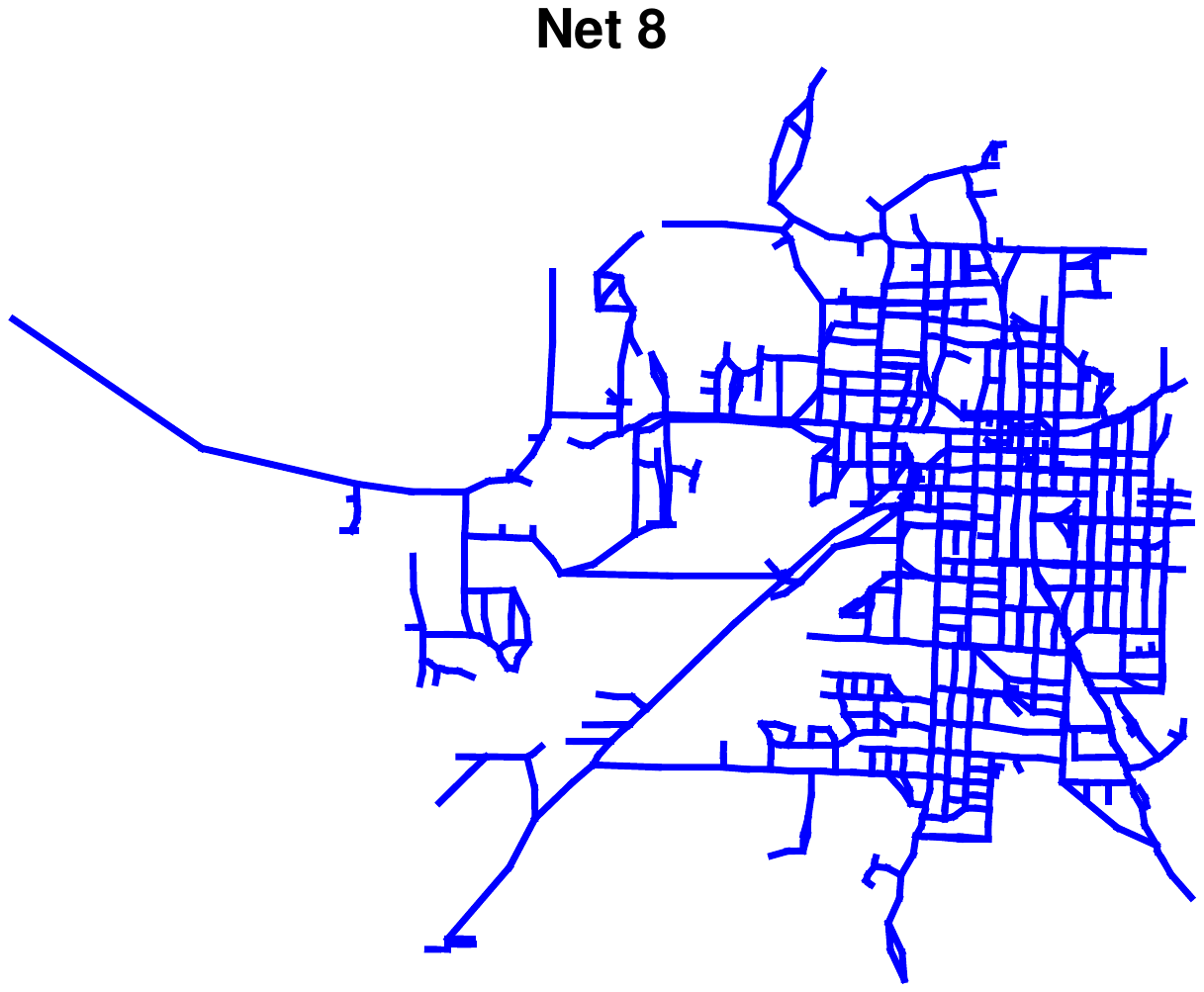}&\includegraphics[trim = 10mm 60mm 10mm 60mm, clip, scale = 0.25]{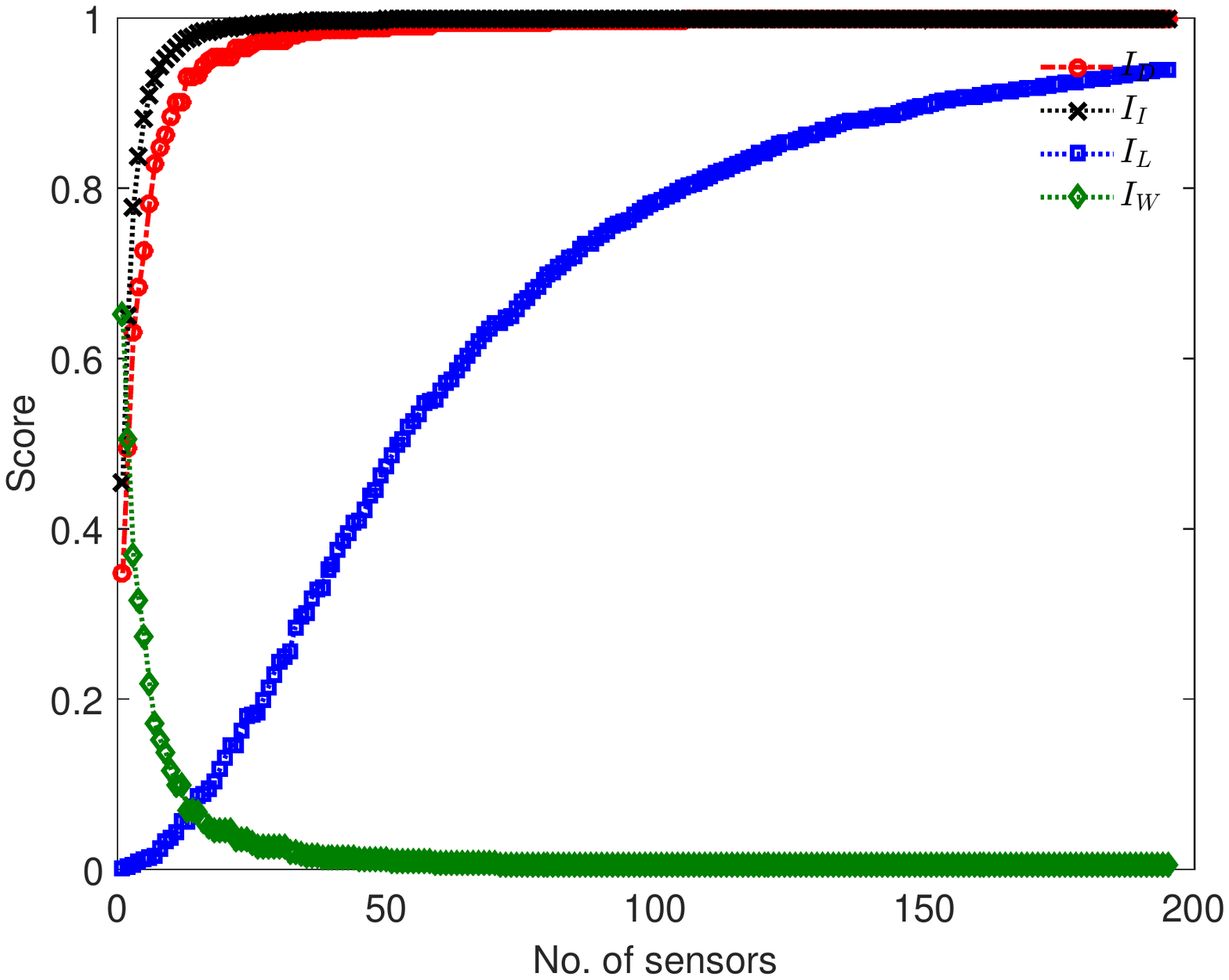}\\
\hline
%
\includegraphics[trim = 10mm 60mm 10mm 60mm, clip, scale = 0.25]{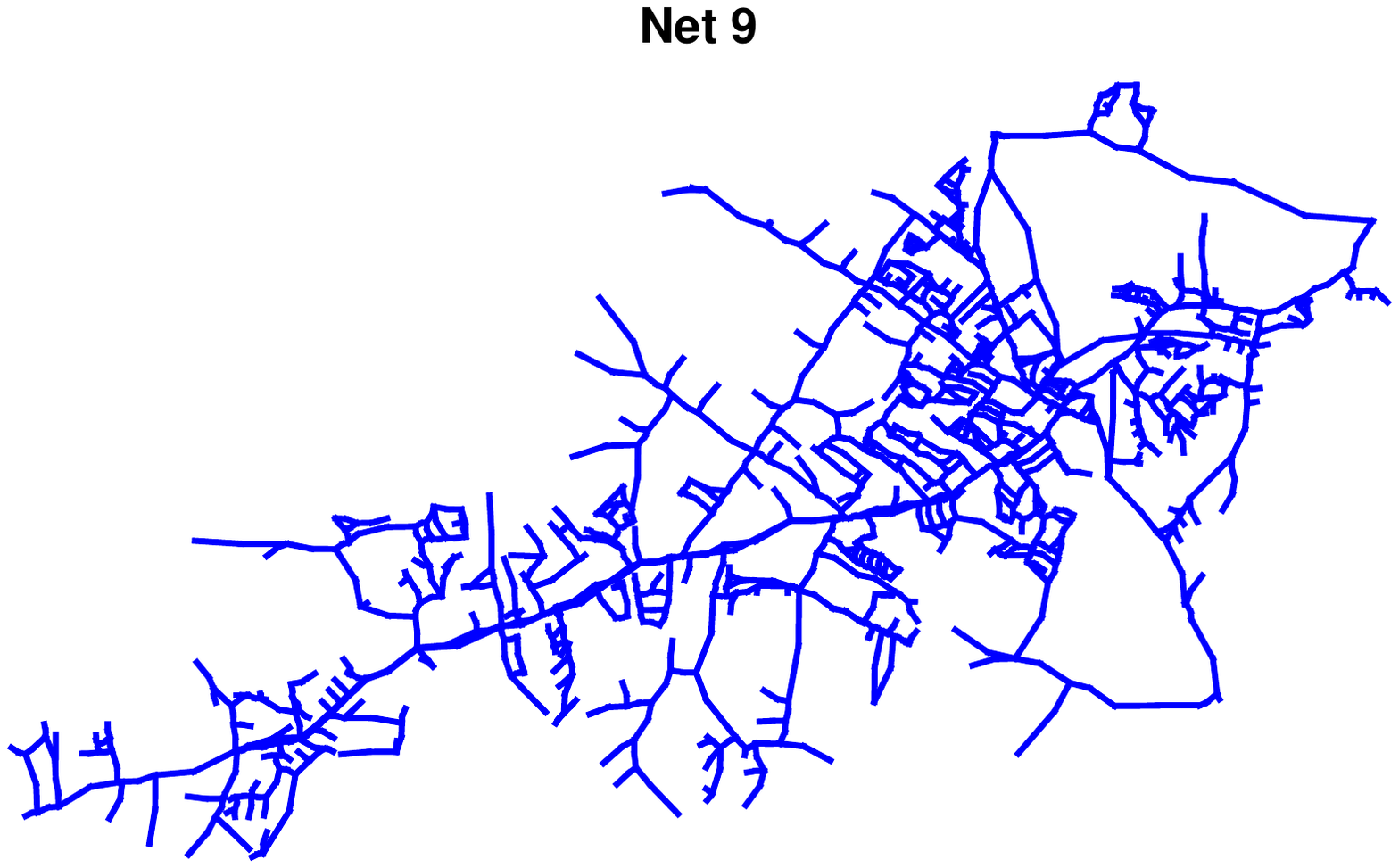}&\includegraphics[trim = 10mm 60mm 10mm 60mm, clip, scale = 0.25]{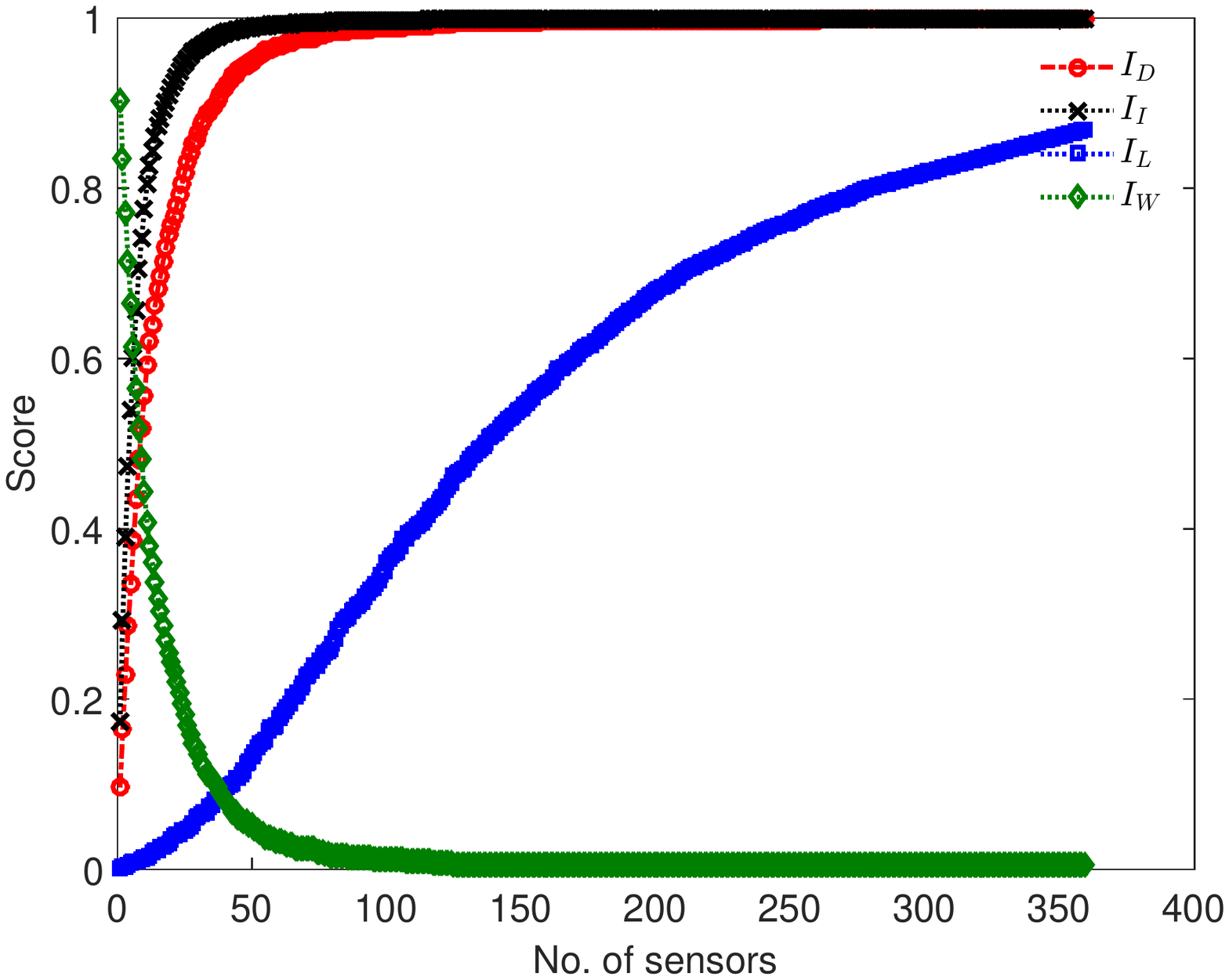}\\
 \hline
\includegraphics[trim = 10mm 60mm 10mm 60mm, clip, scale = 0.25]{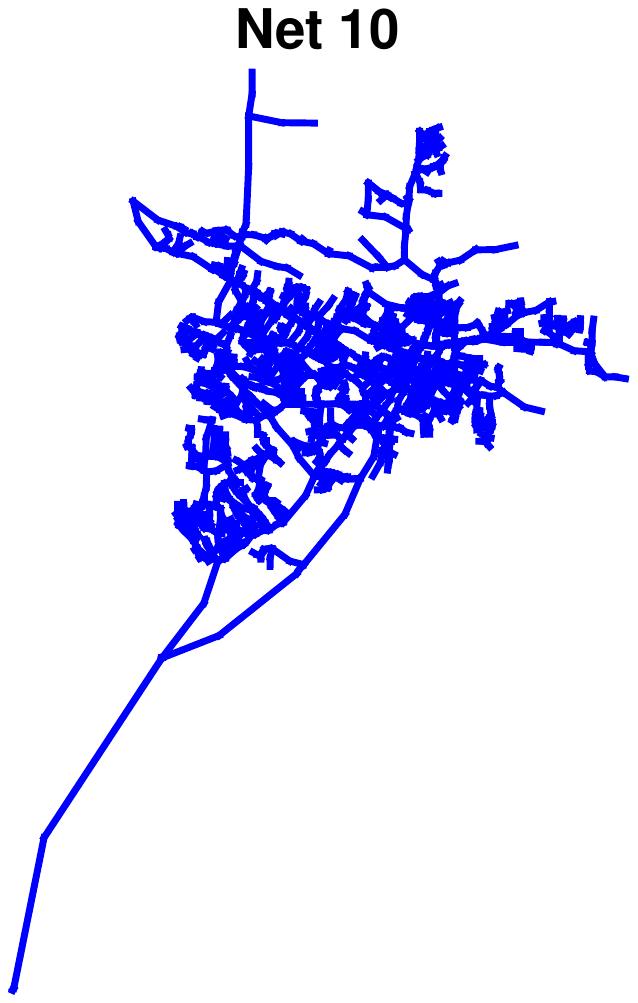}&\includegraphics[trim = 10mm 60mm 10mm 60mm, clip, scale = 0.25]{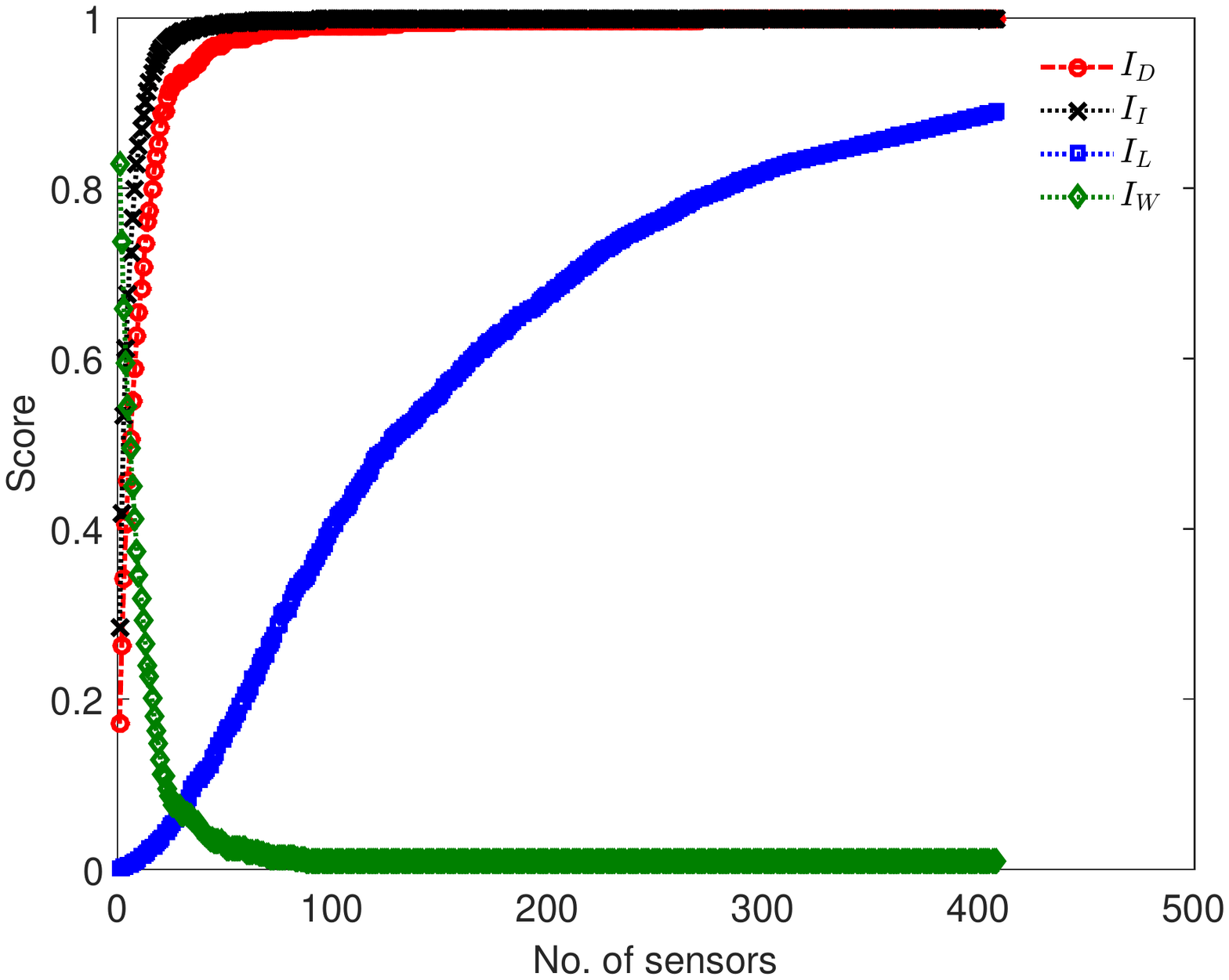}\\
\hline
\end{tabular}
\label{tab:gt}
\end{table*}

\newpage
\begin{table*}[tp]
\caption*{Evaluation on real netowrks}
\centering
\begin{tabular}{|c|c|}
 \hline
\includegraphics[trim = 10mm 60mm 10mm 60mm, clip, scale = 0.25]{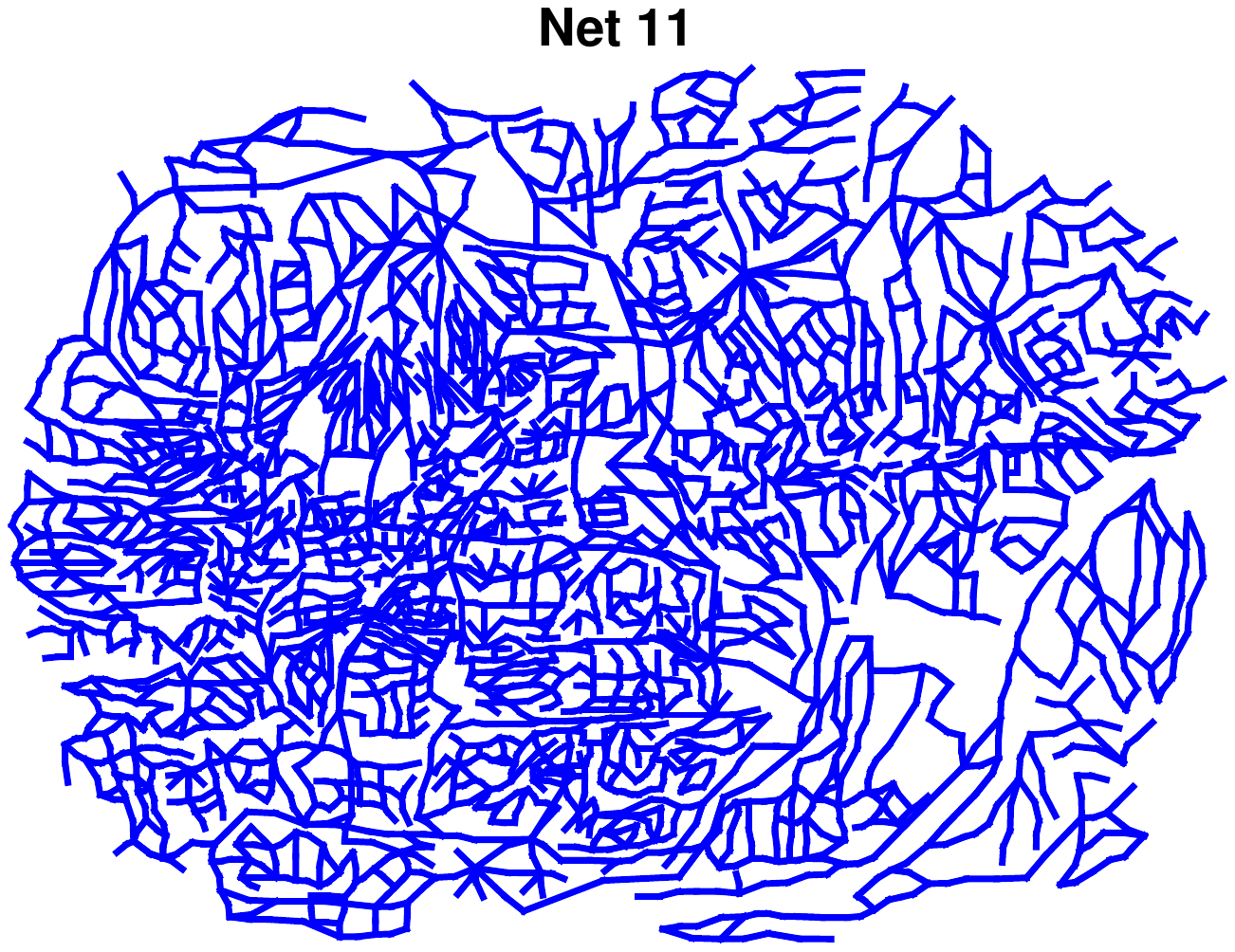}&\includegraphics[trim = 10mm 60mm 10mm 60mm, clip, scale = 0.25]{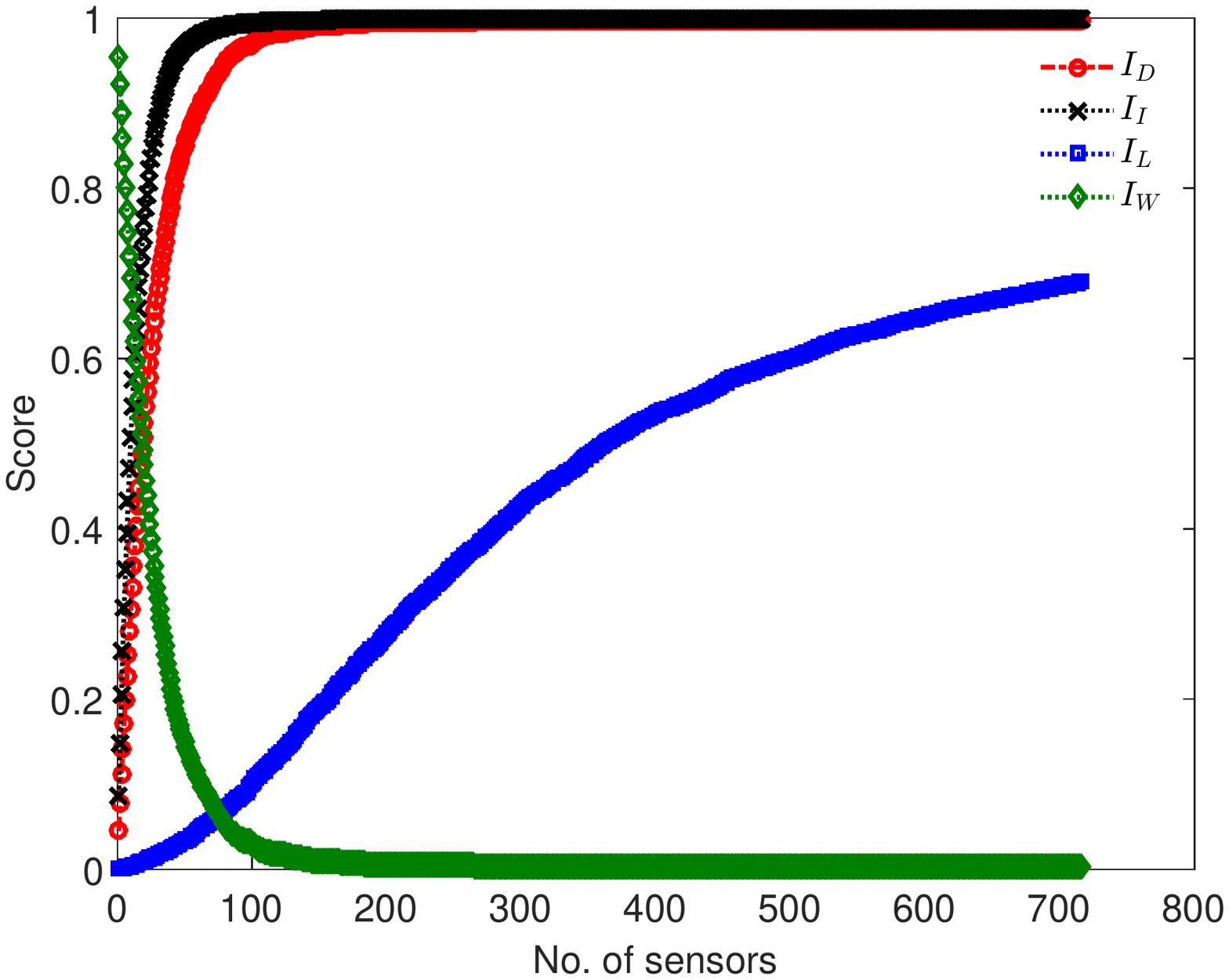}\\
\hline
\includegraphics[trim = 10mm 60mm 10mm 60mm, clip, scale = 0.25]{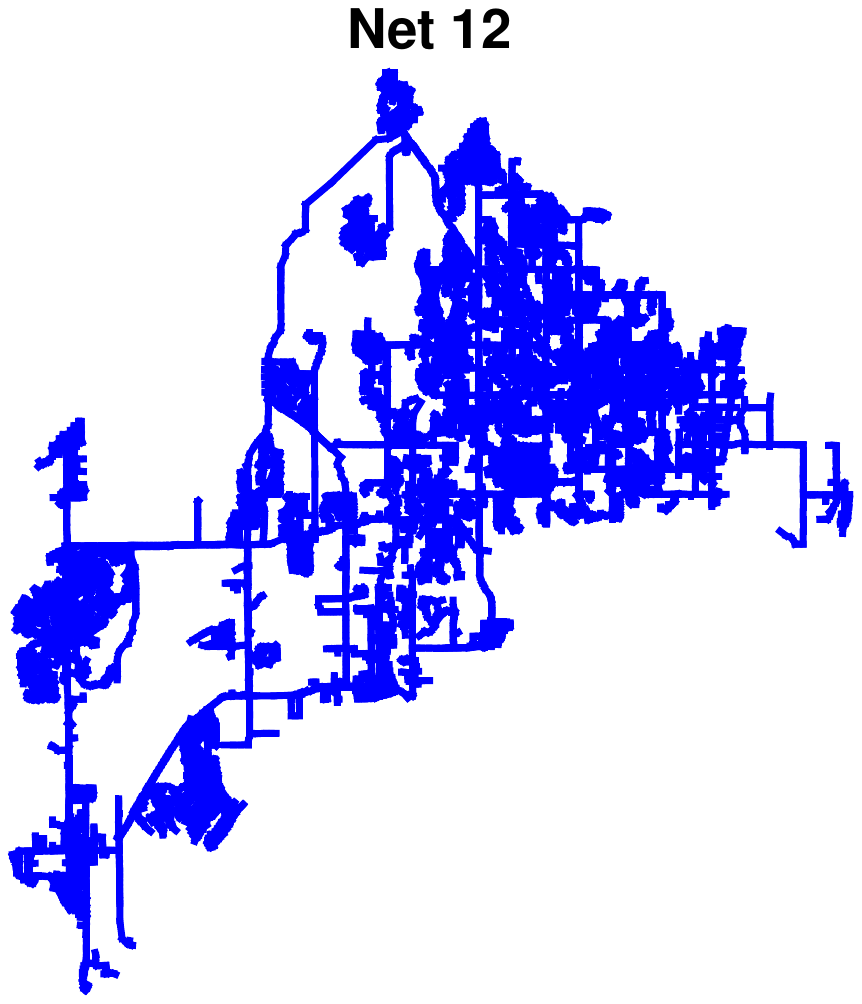}&\includegraphics[trim = 10mm 60mm 10mm 60mm, clip, scale = 0.25]{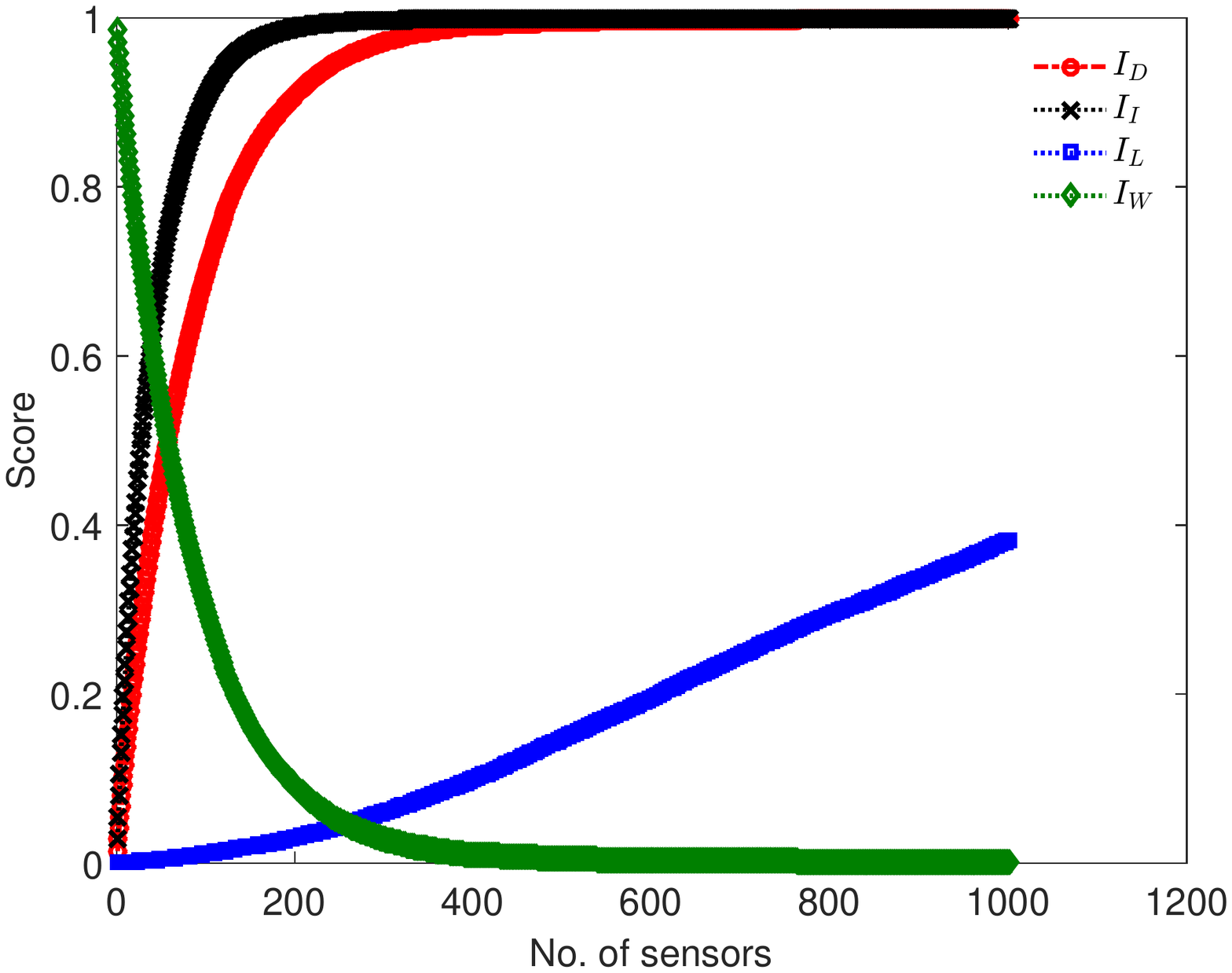}\\
\hline
\end{tabular}
\label{tab:gt}
\end{table*}

\end{document}